\documentstyle[eqsecnum,epsfig,aps,floats,preprint,delarray]{revtex}
 
\def\build#1_#2^#3{\mathrel{
\mathop{\kern 0pt#1}\limits_{#2}^{#3}}}

\newcommand{\be}{\begin{equation}}
\newcommand{\ee}{\end{equation}}
\newcommand{\bea}{\begin{eqnarray}}
\newcommand{\eea}{\end{eqnarray}}

\def\laq{~\raise 0.4ex\hbox{$<$}\kern -0.8em\lower 0.62

ex\hbox{$\sim$}~}

\def\gaq{~\raise 0.4ex\hbox{$>$}\kern -0.7em\lower 0.62
ex\hbox{$\sim$}~}

\newcommand{\whh}{\widehat H}

\begin{document}
\par
\begingroup

\begin{flushright}
IHES/P/02/09 \\ 
Bicocca-FT-02-03 \\
CERN-TH/2002-093
\end{flushright}

{\large\bf\centering\ignorespaces
Violations of the equivalence principle in a dilaton-runaway scenario
\vskip2.5pt}
{\dimen0=-\prevdepth \advance\dimen0 by23pt
\nointerlineskip \rm\centering
\vrule height\dimen0 width0pt\relax\ignorespaces
T. Damour$^{1}$, F. Piazza$^{2}$ and G. Veneziano$^{3,4}$
\par}
{\small\it\centering\ignorespaces
${}^{1}$
Institut des Hautes Etudes Scientifiques, 91440 
Bures-sur-Yvette, France \\
${}^{2}$
Dipartimento di Fisica, Universit\`a di Milano Bicocca \\
Piazza delle Scienze 3, I-20126 Milan, Italy \\
${}^{3}$
Theory Division, CERN, CH-1211 Geneva 23, Switzerland\\
${}^{4}$
Laboratoire de Physique Th\'eorique, Universit\'e Paris
Sud, 91405 Orsay, France\\
\par}

\par
\bgroup
\leftskip=0.10753\textwidth \rightskip\leftskip
\dimen0=-\prevdepth \advance\dimen0 by17.5pt \nointerlineskip
\small\vrule width 0pt height\dimen0 \relax

\begin{abstract}
We explore a version of the cosmological dilaton-fixing and decoupling mechanism  
in which  the dilaton-dependence of the low-energy effective action is extremized
for infinitely large values of the bare string coupling 
$g_s^2 = e^{\phi}$. We study the efficiency with which the dilaton $\phi$ runs away 
towards its ``fixed point'' at infinity during a primordial inflationary stage, and 
thereby approximately decouples from matter. The residual dilaton couplings are found 
to be related to the amplitude of the density fluctuations generated during inflation. 
For the simplest inflationary potential, $V (\chi) = \frac{1}{2} \, m_{\chi}^2 (\phi) 
\, \chi^2$, the residual dilaton couplings are shown to predict violations of the 
universality of gravitational acceleration near the $\Delta a / a \sim 10^{-12}$ level. 
This suggests that a modest improvement in the precision of equivalence principle tests 
might be able to detect the effect of such a runaway dilaton. Under some assumptions
about the coupling of the dilaton to dark matter and/or dark energy, the expected
time-variation of natural ``constants'' (in particular of the fine-structure
constant) might also be large enough to be within reach of improved experimental 
or observational data.
\end{abstract}

 \par\egroup
\thispagestyle{plain}
\endgroup


\section{Introduction}\label{sec1}

All string theory models predict the existence of a scalar partner of the spin 2 
graviton: the dilaton $\phi$, whose vacuum expectation value (VEV) determines the 
string coupling constant $g_s = e^{\phi / 2}$ \cite{Witten}. At tree level,
 the dilaton is massless 
and has gravitational-strength couplings to matter which violate the equivalence 
principle \cite{TV88}. This is in violent conflict with present experimental tests of 
general relativity. It is generally assumed that this conflict is avoided because, 
after supersymmetry breaking, the dilaton might acquire a (large enough) mass (say 
$m_{\phi} \gtrsim 10^{-3} \, {\rm eV}$ so that observable deviations from Einstein's 
gravity are quenched on distances larger than a fraction of a millimeter). However, 
Ref.~\cite{DP94} (see also \cite{DN93}) has proposed a mechanism which can naturally 
reconcile a {\it massless} dilaton with existing experimental data. The basic idea of 
Ref.~\cite{DP94} was to exploit the string-loop modifications of the (four dimensional) 
effective low-energy action (we use the signature $-+++$)
\be
\label{eq1.1}
S = \int d^4 x \sqrt{\widetilde{g}} \left( \frac{B_g (\phi)}{\alpha'} \, \widetilde R + 
\frac{B_{\phi} (\phi)}{\alpha'} \, \lbrack 2 \widetilde{\build\Box_{}^{}} \phi -
(\widetilde{\nabla} \phi)^2 \rbrack - \frac{1}{4} \, B_F 
(\phi) \, \widetilde{F}^2 - V + \cdots \right) \, ,
\ee
i.e. the $\phi$-dependence of the various coefficients $B_i (\phi)$, $i = g , \phi , F, 
\ldots$ , given in the weak-coupling region ($e^{\phi} \to 0$) by series of the form
\be
\label{eq1.2}
B_i (\phi) = e^{-\phi} + c_0^{(i)} + c_1^{(i)} \, e^{\phi} + c_2^{(i)} \, e^{2\phi} + 
\cdots \, ,
\ee
coming from genus expansion of string theory: $B_i = \Sigma_n \, g_s^{2(n-1)} c_n^{(i)}$, 
with $n = 0,1,2,\ldots$. It was shown in \cite{DP94} that, if there exists a special 
value $\phi_m$ of $\phi$ which extremizes all the (relevant) coupling functions 
$B_i^{-1} (\phi)$, the cosmological evolution of the graviton-dilaton-matter system 
naturally drives $\phi$ towards $\phi_m$. This provides a mechanism for fixing a 
massless dilaton at a value where it decouples from matter (``Least Coupling 
Principle''). A simple situation where the existence of a universally extremizing 
dilaton value $\phi_m$ is guaranteed is that of $S$ duality, i.e. a symmetry $g_s 
\leftrightarrow 1/g_s$, or $\phi \rightarrow -\phi$ (so that $\phi_m = 0$).

It has been recently suggested \cite{V01} that the infinite-bare-coupling limit $g_s 
\rightarrow \infty$ ($\phi \rightarrow +\infty$) might yield smooth {\it finite} limits 
for all the coupling functions, namely
\be
\label{eq1.3}
B_i (\phi) = C_i + {\cal O} (e^{-\phi}) \, .
\ee
Under this assumption, the coupling functions are all extremized at infinity, i.e. 
$\phi_m = + \infty$. The late-time cosmology of models satisfying (\ref{eq1.3}) has 
been recently explored \cite{GPV}. In the ``large N''-type toy model of \cite{V01}
it would be natural to expect that the ${\cal O} (e^{-\phi})$ term in Eq.~(\ref{eq1.3})
be {\it positive}, so that $B_i (\phi)$ be {\it minimized} at infinity. This would
correspond to couplings  $\lambda_i (\phi) \sim  B_i^{-1} (\phi) 
= C_i^{-1} - {\cal O} (e^{-\phi})$ which are {\it maximized} at infinity. Note, however,
 that the most relevant cosmological
coupling for this work, the coupling  to the inflaton, $\lambda(\phi)$,
 contained in $V$ (see
 Eq. (\ref{eq2.10}) below) is closer to a $B_i$ than to its inverse. Thus $\lambda(\phi)$
is naturally {\it minimized} at infinity (see further discussion of this point below), a
 crucial property for the attractor mechanism of \cite{DP94,DN93}.

In this paper\footnote{The main results of this work have been recently
summarized in a short note \cite{prl}.} we shall consider in detail the early-time cosmology of models satisfying 
(\ref{eq1.3}). More precisely, our main aims will be: (i) to study the efficiency with 
which a primordial inflationary stage drives $\phi$ towards the ``fixed point'' at 
infinity $\phi_m = + \infty$ (thereby generalizing the work \cite{DV96} which 
considered the inflationary attraction towards a local extremum $\phi_m$), and (ii) to 
give quantitative estimates of the present violations of the equivalence principle (non 
universality of free fall, and variation of ``constants''). Our most important 
conclusion is that the runaway of the dilaton towards strong-coupling (under the 
assumption (\ref{eq1.3})) naturally leads to equivalence-principle violations which are 
rather large, in the sense of not being much smaller than the presently tested 
level $\sim 10^{-12}$. This gives additional motivation for the currently planned 
improved tests of the universality of free fall. Within our scenario, most of the other 
deviations from general relativity (``post-Einsteinian'' effects in gravitationally 
interacting systems: solar system, binary pulsars, ...) are too small to be of 
phenomenological interest. However, under some assumptions about the coupling of $\phi$
to dark matter and/or dark energy, the time variation of the natural ``constants'' 
(notably the fine-structure constant) predicted by our scenario might be large enough
to be within reach of improved experimental and/or observational data. The 
phenomenologically interesting conclusion that equivalence-principle violations are 
generically predicted to be rather large after inflation (in sharp contrast with the 
results of \cite{DV96}) is due to the fact that the attraction towards an extremum at 
infinity is much less effective than the attraction towards a (finite) local extremum 
as originally contemplated in \cite{DP94}. This reduced effectiveness was already 
pointed out in Ref.~\cite{DN93} within the context of equivalence-principle-respecting 
tensor-scalar theories (\`a la Jordan-Fierz-Brans-Dicke).

\section{Dilaton runaway}

In this section we study the dilaton's runaway 
during the various stages of 
cosmological evolution. We first show (subsection
\ref{sec2}) 
that, like in the case of a local extremum \cite{DV96}, inflation is 
particularly efficient in pushing $\phi$ towards the fixed point.
We will then argue (subsection
\ref{sec2b}) that the order of magnitude of the bare string coupling
 $e^{\phi} \simeq e^{c\varphi}$ does not suffer further appreciable changes 
 during all the subsequent evolution.

\subsection{The inflationary period}\label{sec2}

Assuming some primordial inflationary stage driven by the potential energy of an
inflaton 
field $\widetilde{\chi}$, and taking into account generic couplings to the dilaton 
$\phi$, we consider an effective action of the form
\be 
\label{eq2.1}
S = \int d^4 x \sqrt{\widetilde g} \left( \frac{B_g (\phi)}{\alpha'} \, \widetilde{R} + 
\frac{B_{\phi} (\phi)}{\alpha'} \, \lbrack 2 \widetilde{\build\Box_{}^{}} \phi -
(\widetilde{\nabla} \phi)^2 \rbrack
- \frac{1}{2} \, 
B_{\chi} (\phi) (\widetilde{\nabla} \widetilde{\chi})^2 - \widetilde V 
(\widetilde{\chi} , \phi) \right) \, .
\ee
In this string-frame action, the dilaton dependence of all the functions $B_i (\phi)$, 
$\widetilde V (\widetilde{\chi} , \phi)$ is assumed to be of the form (\ref{eq1.2}). It 
is convenient to replace the ($\sigma$-model) string metric $\widetilde{g}_{\mu \nu}$ 
by the conformally related Einstein metric $g_{\mu \nu} = C \, B_g (\phi) \, 
\widetilde{g}_{\mu \nu}$, and the dilaton field by the variable
\be
\label{eq2.2}
\varphi = \int d \phi \left[ \frac{3}{4} \left( \frac{B'_g}{B_g} \right)^2 + 
\frac{B'_{\phi}}{B_g} + \frac{1}{2} \, \frac{B_{\phi}}{B_g} \right]^{\frac{1}{2}} \, , \;
B' \equiv \partial B/\partial \phi \; .
\ee
The normalization constant $C$ is chosen so that the string units coincide with the 
Einstein units when $\phi \rightarrow + \infty$: $C \, B_g (+\infty) = 1$. [Note that 
$C = 1/C_g$ in terms of the general notation of Eq.~(\ref{eq1.3}).] Introducing the 
(modified) Planck mass
\be
\label{eq2.3}
\widetilde{m}_P^2 = \frac{1}{4\pi G} = \frac{4}{C \alpha'}\, ,
\ee
and replacing also the inflaton by the dimensionless variable $\chi = C^{-1/2} \, 
\widetilde{m}_P^{-1} \, \widetilde{\chi}$, we end up with an action of the form
\be
\label{eq2.4}
S = \int d^4 x \sqrt{g} \left[ \frac{\widetilde{m}_P^2}{4} \, R - 
\frac{\widetilde{m}_P^2}{2} \, (\nabla \varphi)^2 - \frac{\widetilde{m}_P^2}{2} \, F 
(\varphi) (\nabla \chi)^2 - \widetilde{m}_P^4 \, V (\chi , \varphi) \right] \, ,
\ee
where
\be
\label{eq2.5}
F(\varphi) = B_{\chi} (\phi) / B_g (\phi) \, , \ V(\chi , \varphi) = C^{-2} \, 
\widetilde{m}_P^{-4} \, B_g^{-2} (\phi) \, \widetilde V (\widetilde{\chi} , \phi) \, .
\ee
In view of our basic assumption (\ref{eq1.3}), note that, in the strong-coupling limit 
$\phi \rightarrow +\infty$, $d\varphi / d \phi$ tends, according to Eq.~(\ref{eq2.2}), 
to the constant $(C_{\phi} / 2 C_g)^{1/2}$, while the dilaton-dependent factor 
$F(\varphi)$ in front of the inflaton kinetic term tends to the constant $C_{\chi} / 
C_g$. The toy model of Ref.~\cite{V01} suggests that the various (positive) constants $C_i$ in 
Eq.~(\ref{eq1.3}) are all largish and comparable to each other. We shall therefore 
assume that the various ratios $C_i / C_j$ are of order unity. The most important such 
ratio for the following is $c \equiv (2 C_g / C_{\phi})^{1/2}$ which gives the 
asymptotic behaviour of the bare string coupling as
\be
\label{eq2.6}
g_s^2 = e^{\phi} \simeq e^{c\varphi} \, .
\ee
In view of the fact that, in the strong-coupling limit we are interested in, the factor 
$F(\varphi)$ in Eq.~(\ref{eq2.4}) quickly tends to a constant, we can simplify our 
analysis (without modifying the essential physics) by replacing it by a constant (which 
can then be absorbed in a redefinition of $\chi$). Henceforth, we shall simply take 
$F(\varphi) = 1$. [See, however, the comments below concerning the self-regenerating
inflationary regime.]

Following \cite{DN93,D95} it is then useful to combine the Friedmann equations for the 
scale factor $a(t)$ during inflation ($ds^2 = -dt^2 + a^2 (t) \, \delta_{ij} \, dx^i \, 
dx^j$) with the equations of motion of the two scalar fields $\chi (t)$, $\varphi (t)$, 
to write an autonomous equation describing the evolution of the two scalars in terms of 
the parameter
\be
\label{eq2.7}
p = \int H dt = \int \frac{\dot a}{a} \, dt = \ln \, a + {\rm const}
\ee
measuring the number of $e$-folds of the expansion. For any multiplet of scalar fields, 
$\mbox{\boldmath$\varphi$} = (\varphi^a)$, this yields the simple equation 
\cite{DN93,D95}
\be
\label{eq2.8}
\frac{2}{3 - \mbox{\boldmath$\varphi$}'^2} \, \mbox{\boldmath$\varphi$}'' + 2 
\, \mbox{\boldmath$\varphi$}' = - \nabla_{\mbox{\boldmath$\varphi$}} \ln \vert V 
(\mbox{\boldmath$\varphi$}) \vert \, ,
\ee
where $\mbox{\boldmath$\varphi$}' \equiv d \mbox{\boldmath$\varphi$} / dp$, and where 
all operations on $\mbox{\boldmath$\varphi$}$ are covariantly defined in terms of the 
$\sigma$-model metric defining the scalar kinetic terms ($d\sigma^2 = \gamma_{ab} 
(\varphi) \, d\varphi^a \, d\varphi^b$). In our simple model (with $F(\varphi) = 1$), 
we have a flat metric $d\sigma^2 = d \varphi^2 + d\chi^2$. [Note that, when 
$\gamma_{ab} (\varphi)$ is curved the acceleration term $\mbox{\boldmath$\varphi$}''$ 
involves a covariant derivative.]

The generic solution of Eq.~(\ref{eq2.8}) is easily grasped if one interprets it as a 
mechanical model: a particle with position $\mbox{\boldmath$\varphi$}$, and 
velocity-dependent mass $m(\mbox{\boldmath$\varphi$}') = 
2/(3-\mbox{\boldmath$\varphi$}'^2)$, moves, in the ``time'' $p = \ln a + {\rm cst}$, in 
the manifold $d \sigma^2$ under the influence of an external potential $\ln \vert V 
(\mbox{\boldmath$\varphi$})\vert$ and a constant friction force $-2 \, 
\mbox{\boldmath$\varphi$}'$. If the curvature of the effective potential $\ln \vert V 
(\mbox{\boldmath$\varphi$})\vert$ is sufficiently small the motion of 
$\mbox{\boldmath$\varphi$}$ rapidly becomes slow and friction-dominated:
\be
\label{eq2.9}
2 \, \frac{d\mbox{\boldmath$\varphi$}}{dp} \simeq - \nabla_{\mbox{\boldmath$\varphi$}} 
\ln V (\mbox{\boldmath$\varphi$}) \, .
\ee
Eq.~(\ref{eq2.9}) is equivalent to the usual ``slow roll'' approximation.

 Consistently with our general assumption (\ref{eq1.3}), we consider potentials allowing
a strong-coupling expansion of the form:
\be
\label{genpot}
V (\chi , \varphi) = V_0(\chi) + V_1(\chi) e^{-c\varphi} + {\cal O}(e^{-2c\varphi}) \, ,
\ee
where $V_0(\chi)$ is a typical chaotic-inflation potential with 
$V_0(0)=0$, while $V_1(0) = v_1 \ge 0$ can possibly provide (if $v_1 >0$)
 the effective cosmological constant
driving   today's acceleration in the scenario of \cite{GPV}.
For the sake of simplicity we shall discuss mainly the ``factorized'' 
power-law case $V_0(\chi) \sim V_1(\chi) \sim \chi^n$ for which we can
conveniently write $V$ in the form:
\be
\label{eq2.10}
V (\chi , \varphi) = \lambda (\varphi) \, \frac{\chi^n}{n} \, ,
\ee
 with a dilaton-dependent coupling constant $\lambda (\varphi)$  of the form
\be
\label{eq2.11}
\lambda (\varphi) = \lambda_{\infty} (1 + b_{\lambda} \, e^{-c\varphi}) \, .
\ee
This example belongs to the class of the
 two-field inflationary potentials discussed in \cite{L90}. We have checked that
our results remain qualitatively the same for the more general potential (\ref{genpot})
provided that  $V_0(\chi)$ and $V_1(\chi)$ are not extremely diffent and given
the fact that $v_1$ is phenomenologically constrained to be very small.
[Note that, within the simplified model (\ref{eq2.11}), the ratio  $V_1(\chi)/V_0(\chi)$
is equal to the constant coefficient  $b_{\lambda}$.]

The universal (positive) constant $c$ appearing in the exponential $e^{-c\varphi}$
  is the same as in 
Eq.~(\ref{eq2.6}) [i.e. $c \equiv (2 C_g / C_{\phi})^{1/2}$, which is expected to be of 
order unity]. The coefficient $b_{\lambda}$ in (\ref{eq2.11}) is such that $b_{\lambda} 
\, e^{-c \varphi} \simeq b_{\lambda} \, e^{-\phi}$ roughly corresponds to a combination 
of terms $\sim \pm \, C_i^{-1} \, {\cal O} (e^{-\phi})$ coming from the strong-coupling 
asymptotics of several $B_i (\phi)$, Eq.~(\ref{eq1.3}) (see Eq.~(\ref{eq2.5})). In the 
toy model of \cite{V01} one would therefore expect $b_{\lambda}$ to be smallish. 
Anyway, we shall see that in final results only the ratios of such $b_i$ coefficients 
enter. More important than the magnitude of $b_{\lambda}$ is its sign. It is crucial 
for the present strong-coupling attractor scenario to assume that $b_{\lambda} > 0$, 
i.e that $\lambda (\varphi)$ reaches a {\it minimum} at strong-coupling, $\varphi 
\rightarrow + \infty$. Note again that this behaviour is consistent with the
simple ``large $N$''-type idea of \cite{V01} if we assimilate $\lambda (\varphi)$ to one
of the inverse couplings $B_i$ appearing in (\ref{eq1.1})
 (for instance $B_F \sim g_F^{-2}$, where $g_F$ is a 
gauge coupling), rather than to
 the coupling itself.
If the latter were the case, $\lambda (\varphi)$ would reach 
a {\it maximum} as $\phi \rightarrow + \infty$, and the attractor mechanism of 
\cite{DP94} would drive $\phi$ towards weak coupling ($\phi \rightarrow - \infty$).
However, the Einstein-frame $\phi$-dependence of $V(\chi)$ gets contributions from 
several $B_i^{\pm n}(\phi)$, Eq.~(\ref{eq2.5}), which might conspire to minimize it
at strong coupling. This feature is also probably necessary in order to solve
the cosmological-constant problem through some argument by which the vacuum at infinity
has vanishing energy density.

Substituting the potential (\ref{genpot}) into the slow roll equation (\ref{eq2.9})
and assuming (for simplicity) that $ V_1(\chi) e^{-c\varphi}$ is significantly smaller 
than $V_0(\chi)$ 
leads to a decoupled set of evolution equations for $\chi$ and $\varphi$ (where 
$V' \equiv \partial V / \partial \chi$):
\be
\label{eq2.12}
\frac{d \chi}{dp} = - \frac{1}{2} \, \frac{V'_0}{V_0} \, ,
\ee
\be
\label{eq2.13}
\frac{d \varphi}{dp} =  \frac{1}{2}  \, c \, e^{-c \varphi} \frac{V_1}{V_0} \, .
\ee
Given some ``initial'' conditions $\chi_{\rm in}$, $\varphi_{\rm in}$ (discussed below)
at some starting 
point, say $p=0$, the solution of Eqs.~(\ref{eq2.12}), (\ref{eq2.13}) is simply

\be
\label{eq2.14}
p =  \, 2 \int_{\chi}^{\chi_{\rm in}} d\bar\chi \,
\bar\chi \left( \frac{V_0(\bar\chi)}{\bar\chi \, V'_0(\bar\chi)}\right) \, ,
\ee
\be
\label{eq2.15}
e^{c \varphi} = e^{c \varphi_{\rm in}} +   \, \frac{c^2}{2} \, 
\int dp \, \frac{V_1(\chi (p))}{V_0(\chi (p))} \, ,
\ee
which simply become:
\be
\label{eq2.16}
p =  \frac{1}{n} \left( \chi_{\rm in}^2 - \chi^2 \right)  \, , \quad 
e^{c \varphi} + \frac{b_{\lambda} \, c^2}{2n} \, \chi^2 = {\rm const}. = e^{c 
\varphi_{\rm in}} + \frac{b_{\lambda} \, c^2}{2n} \, \chi_{\rm in}^2 \, ,
\ee
in the simplified case of eqs. (\ref{eq2.10}), (\ref{eq2.11}).

Equations (\ref{eq2.16}) show that, in order for the string coupling 
 $g_s^2 \simeq e^{c \varphi}$ to have reached large values at the end of inflation,
a large total number of e-folds must have occurred while the
(dimensionless) inflaton field $\chi$ decreases from a large initial value, to a
value of order unity (in Planck units). To get a quantitative estimate of the string coupling
at the end of inflation we need to choose the initial conditions 
$\chi_{\rm in}$, $\varphi_{\rm in}$. A physically reasonable way (which is further
discussed below) of choosing $\chi_{\rm in}$
is to start the classical evolution (\ref{eq2.12})-(\ref{eq2.16}) at the exit of the
era of self-regenerating inflation (see \cite{L90} and references therein).
We will now show how to relate the exit from self-regenerating inflation 
 to the size of density fluctuations generated by inflation.

 Let us recall (see \cite{L90} and references 
therein) that the density fluctuation $\delta \equiv \delta \rho / \rho$ on large scales 
(estimated in the one-field approximation where the inflaton $\chi$ is the main 
contributor) is obtained by evaluating the expression
\be
\label{eq2.19}
\delta (\chi) \simeq \frac{4}{3} \, \frac{1}{\pi} \left( \frac{2}{3} 
\right)^{\frac{1}{2}} \, \frac{V^{3/2}}{\partial_\chi V}
\ee
at the value $\chi = \chi_{\times}$, at which the physical scale
we are considering crossed the horizon outwards during inflation. For the scale
corresponding to  our present
horizon this usually corresponds to  a value  $\chi_{\times}(H_0)$  ($\chi_H$ for short)  
reached some $ 60$ $e$-folds before 
the end of slow-roll. From Ref.~\cite{L90}, $\chi_H \simeq 5 \sqrt n$ for the model 
(\ref{eq2.10}) (and with our modified definition of $\chi$). 
The numerical value of $\delta_H \equiv \delta (\chi_H)$ which 
is compatible with cosmological data (structure 
formation and cosmic microwave background) is $\delta_H \simeq 5 \times 10^{-5}$.
 In the model 
(\ref{eq2.10}) the function $\delta (\chi)$ defined by (\ref{eq2.19}) scales with 
$\chi$ as $\chi^{\frac{n+2}{2}}$. Putting together this information we obtain a
relation between  $\chi_{\rm in}$ and $\delta(\chi_{\rm in})$, which involves
the value of the observable horizon-size fluctuations 
$\delta_H \equiv \delta (\chi_H)$:
\be
\label{eq2.21}
\frac{\delta (\chi_{\rm in})}{\delta (\chi_H)} = \left( \frac{\chi_{\rm in}}{\chi_H} 
\right)^{\frac{n+2}{2}} ,
\ee
i.e.
\be
\label{eq2.22}
\chi_{\rm in} \simeq \chi_H \, \left(\frac{\delta_{\rm in}}{\delta_H} \right)^{ \frac{2}{n+2}} 
\simeq \, 5 \sqrt n \, 
\left(\frac{\delta_{\rm in}}{\delta_H} \right)^{ \frac{2}{n+2}} \, ,
\ee
where we introduced the short-hand notation 
$\delta_{\rm in} \equiv \delta (\chi_{\rm in})$.

Inserting Eq.~(\ref{eq2.22}) into Eq.~(\ref{eq2.15}) we then obtain the following 
estimate of the string coupling constant after inflation as a function of 
$\varphi_{\rm in}$ and $\delta(\chi_{\rm in})$
\be
\label{eq2.23}
e^{c \varphi_{\rm end}} -  e^{c \varphi_{\rm in}} \, \simeq \, \frac{c^2}{2}
\langle V_1/V_0 \rangle  p \, \sim \, \frac{c^2}{2n} \langle V_1/V_0 \rangle  
\chi_{\rm in}^2 \, \sim \, 
\frac{25 c^2}{2} \, \langle V_1/V_0 \rangle  \, 
\left(\frac{\delta_{\rm in}}{\delta_H} \right)^{ \frac{4}{n+2}} \, ,
\ee
where $\langle V_1/V_0 \rangle $ denotes the average value of $V_1/V_0$: $\langle V_1/V_0 \rangle  \equiv
 \int dp (V_1/V_0)/\int dp $ [note that this average ratio is equal to $b_{\lambda}$
 in the simplified model (\ref{eq2.11})].
 
To get a quantitative estimate of $e^{c \varphi_{\rm end}}$ we still need to 
estimate the value of $\delta (\chi_{\rm in})$ corresponding to the chosen 
``initial'' value of the inflaton.  As we will now check,
taking for $\chi_{\rm in}$ the value corresponding to the exit from self-regenerating 
inflation corresponds simply to taking $\delta (\chi_{\rm in}) \sim 1$. Indeed, let us
first recall that, during inflation, each (canonically normalized) scalar field
(of mass smaller than the expansion rate $H$) undergoes typical quantum fluctuations
of order $H/(2 \pi)$, per Hubble time (see, e.g., \cite{L90}).
 This implies (for our dimensionless fields) that the value of $\chi$ at the exit
 from self-regeneration, say $\chi_{\rm ex}$, is characterized by
 ${\widehat H}_{\rm ex}/(2 \pi) \approx \lbrack \partial_\chi V/(2 V) \rbrack_{\rm ex}$,
 where ${\widehat H} \equiv H / \widetilde{m}_P $ is the dimensionless Hubble expansion 
rate and where the right-hand side (RHS) is the classical change of $\chi$ per Hubble time
 (corresponding to the RHS of Eq.(\ref{eq2.12})). Using Friedmann's equation 
 (in the slow-roll approximation) ${\widehat H}^2_{\rm ex} \approx
 (2/3) V(\chi_{\rm ex})$, it is easily seen that that the exit from
 self-regeneration corresponds to $\delta (\chi_{\rm ex}) \approx 4/3 \sim 1$.
 It is, a posteriori, physically quite reasonable to start using the classical evolution
system only when the (formal extrapolation) of the density fluctuation $\delta (\chi)$ 
becomes smaller than one.

Within some approximation (see \cite{L90}), one can implement the effect of the 
combined quantum fluctuations of $(\varphi, \chi)$ 
 by adding  random terms with r.m.s. values 
$\widehat H/2\pi$  on the right hand side of equations (\ref{eq2.12}) and (\ref{eq2.13}),
$d \chi/dp$ and $d \varphi/dp$ being precisely the shifts of the fields in a Hubble time.
The system of equations becomes thus of the Langevin-type 
\begin{equation}
\label{eq4.1}
\frac{d \chi}{d p} \, = 
\, - \frac{1}{2} \, \frac{V'_0}{V_0}\,  +\, \frac{\whh}{2\pi}\, \xi_1 \, ,
\end{equation}
\begin{equation}
\label{eq4.2}
\frac{d \varphi}{dp} \, = \, \frac{1}{2} \, c \, e^{-c \varphi} \frac{V_1}{V_0}\,  +\, \frac{\whh}{2\pi} \,
\xi_2 \, ,
\end{equation} 
where ${\widehat H} \approx [(2/3) V(\chi,\phi)]^{1/2}$ (in the slow-roll approximation) is the
dimensionless expansion rate, and where $\xi_1$ and $\xi_2$ are
(independent) normalized random white noises:
\be
\langle\xi_i(p_1)\, \xi_j(p_2) \rangle \, \, =\, \delta_{ij}\delta(p_1-p_2), \qquad i,j = 1,2 \, .
\ee

When the random force terms dominate the evolution
in either equation (\ref{eq4.1}) or equation (\ref{eq4.2}) the quasi-classical description 
(\ref{eq2.12}) , (\ref{eq2.13}) breaks down. 
The phase space of the sytem can thus be roughly divided into four regions 
according to whether the evolution of none, one or both of the two fields
 is dominated by quantum fluctuations. 
This is depicted in Figure 1 where such regions are delimited by dashed, 
thick curves in the case of a power-law potential (\ref{eq2.10}). 

\begin{figure}[t] \label{figure1}
\begin{center}
\includegraphics[height=9cm]{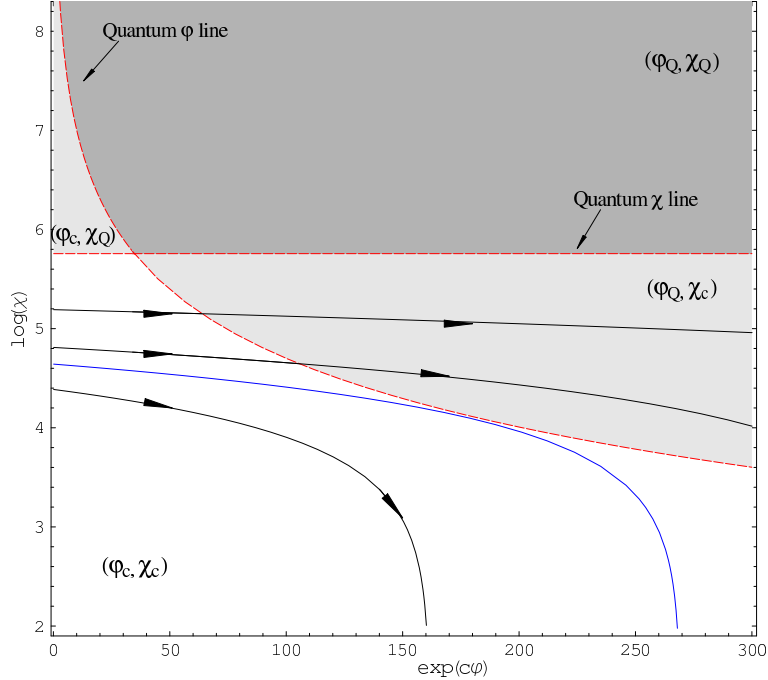}
\vskip 3mm
\caption{\sl The phase space of the system is represented in the case of  
a power--law potential (\ref{eq2.10}) with $n=2$, $b_\lambda = 0.1$ and 
$\lambda_\infty = 10^{-10}$. The 
thick--dashed (red) curves delimitate the quantum behaviour of the two fields, 
the horizontal curve $\chi = \lambda_\infty^{-1/(n+2)} $  and the 
hyperbola-like curve $\chi = b_\lambda^{2/n} \lambda_\infty^{-1/n} e^{-2 c\varphi/n}$
being the limit of the quantum behaviour for $\chi$ and $\varphi$ respectively. 
In the white region both fields have a classical behaviour. The last 
``fully classical'' trajectory has been represented by a thick (blue) curve.
The bright--gray regions are those where either the $\varphi$ or the $\chi$ evolution
are dominated by quantum fluctuations. The fully--quantum region 
is the dark--gray region on the top right.}
\end{center}
\end{figure}

Apart from factors of order one, the evolution of the inflaton $\chi$ is quasi-classical
in the region under the line $\chi = \lambda_\infty^{-1/(n+2)}$. 
In the chaotic inflationary models \cite{L90} 
such an inflaton's value corresponds to
the exit from the self-regenerating regime and to the beginning of the quasi-classical
slow-roll inflation. As mentioned above it 
also corresponds to a perturbation $\delta(\chi) \sim 1$.
Somewhat surprisingly, in the model at hand, however, the quasi-classical region for the 
inflaton evolution 
$\chi \lesssim \lambda_\infty^{-1/(n+2)}$ is  affected
by quantum fluctuations that still dominate the  evolution of $\varphi$
in  the region above the hyperbola-like 
curve $\chi = b_\lambda^{2/n} \lambda_\infty^{-1/n} e^{-2 c\varphi/n}$.
We must therefore study, in some detail, the evolution of the system 
in the presence of the noise term for $\varphi$ as in eq. (\ref{eq4.2}), assuming 
a quasi classical evolution for $\chi$. This is done in the appendix. 
The final result is that, if we start at $\chi \lesssim \lambda_\infty^{-1/(n+2)}$, 
(classical evolution for $\chi$)  
the average value of $e^{c\varphi}$ 
is {\it multiplicatively renormalized}, by a factor of order unity, with respect to 
 the classical trajectory $e^{c\varphi_{\rm cl}}$, given by solving equations
 (\ref{eq2.12}) - (\ref{eq2.13}), i.e.
$\langle e^{c\varphi} \rangle = {\cal O}(1) e^{c\varphi_{\rm cl}}$. One also finds
that the dispersion of $e^{c\varphi}$ around its average value is comparable to its
average value.

We shall not try to discuss here what happens in the self-regenerating region
$\chi \agt \chi_{\rm in} \sim \lambda_\infty^{-1/(n+2)}$. Let us recall that the simple decoupled
system (\ref{eq2.12}) , (\ref{eq2.13}) was obtained by neglecting the kinetic coupling term
$F(\varphi)$ in Eq.~(\ref{eq2.4}). If we were to consider a more general model, we would have
more coupling between $\chi$ and $\varphi$ and we would expect that (contrary to Fig. 1 which
exhibits a ``classical $\varphi$ region'' above the ``quantum $\chi$ line'') the evolution
in the self-regenerating region involves a strongly coupled system of Langevin equations.
Then, as discussed in \cite{L90}, solving such a system necessitates to give boundary conditions
on all the boundaries of the problem: notably for $\chi \to \infty$, but also for 
$\varphi \to +\infty$ and $\varphi \to -\infty$. We leave to future work such an
investigation (and a discussion of what are reasonable boundary conditions). In this work
we shall content ourselves with ``starting'' the evolution on the quantum $\chi$ boundary
line $\chi_{\rm in}$ with some value $\varphi = \varphi_{\rm in}$, assuming that
$ e^{c\varphi_{\rm in}}$ is smaller than the driving effect due to inflation, i.e. than
the RHS of Eq.~(\ref{eq2.23}). [This assumption is most natural in a work aimed at 
studying the ``attracting'' effect due to primordial inflation.]

Going back to our result (\ref{eq2.23}), we can now insert, 
according to the preceding discussion, the values $\delta_{\rm in}=1$ and 
$e^{c \varphi_{\rm in}} \ll e^{c \varphi_{\rm end}}$.
Finally, in the simplified model (\ref{eq2.10}), (\ref{eq2.11}), we get the estimate:
 \be
\label{eq2.23'}
e^{c \varphi_{\rm end}}  = {\cal O}(1) \cdot e^{c\varphi_{\rm cl,\, end}} \sim
{\cal O}(1) \cdot \frac{25 c^2}{2} \, b_{\lambda} \, 
\left(\delta_H\right)^{ -\frac{4}{n+2}} \, .
\ee

A more general analysis, based on the potential (\ref{genpot})
 leads to the same final result but with $n$ replaced
 by some average value of $\frac{\chi V_{0,\chi}}{V_0}$, and with $b_{\lambda}$ replaced by
 some average of the ratio $V_1/V_0$.
 Note that 
smaller values of the exponent $n$ lead to larger values of $e^{c \varphi_{\rm end}}$, i.e. 
to a more effective attraction towards the ``fixed point at infinity''. The same is true
if we take different exponents $n_0$ and $n_1$ (for $V_0$ and $V_1$ respectively) 
and assume $V_1(\chi_{\rm in})
 \gg V_0 (\chi_{\rm in})$ to hold as a result of $\chi_{\rm in} \gg 1$ and 
$n_1 > n_0$. Also note that, 
numerically, if we consider $n=2$, i.e. the simplest chaotic-inflation potential $V = 
\frac{1}{2} \, m_{\chi}^2 (\varphi) \, \chi^2$, Eq.~(\ref{eq2.23'}) involves the large 
number $12.5 \times \delta_H^{-1} \sim 2.5 \times 10^5$. In the case where $n=4$, i.e. $V = 
\frac{1}{4} \, \lambda (\varphi) \, \chi^4$, we have instead the number $12.5 \times 
\delta_H^{-2/3} \sim 0.92 \times 10^4$. To understand the phenomenological meaning of these 
numbers we need to relate $e^{c \varphi_{\rm end}}$ to the present, observable deviations 
from general relativity. This issue is addressed in Section \ref{sec3} after having argued
that the post-inflationary evolution of $\varphi$ is sub-dominant.

\subsection{Attraction of $\varphi$ by the subsequent cosmological evolution}\label{sec2b}

We have discussed above the efficiency with which inflation drives the dilaton towards a 
fixed point at infinity. We need to complete this discussion by estimating the effect 
of the many $e$-folds of expansion that took place between the end of inflation and the 
present time. To address this question, we need to study in more detail the 
coupling of a runaway dilaton to various types of matter, say a multi-component 
distribution of (relativistic or non-relativistic) particles. 
We work in the Einstein frame, with an action of the type

\begin{eqnarray}
S = \int d^4x \sqrt{g}\left[\frac{\widetilde{m}_P^2}{4}R-\frac{\widetilde{m}_P^2}{2}
(\nabla \varphi)^2  - \frac{1}{4} B_F(\varphi) F^2 + \dots \right]  
\qquad \qquad\\[2mm]
 \qquad   \qquad  \qquad  -  \sum_A \int m_A[\varphi(x_A)] \sqrt{-g_{\mu \nu}(x_A) dx_A^\mu dx_A^\nu}\, . 
\nonumber
\end{eqnarray}
Following \cite{TV88,DP94}, one 
introduces the crucial dimensionless quantity
\be
\label{eq3.1}
\alpha_A (\varphi) \equiv \frac{\partial \ln m_A (\varphi)}{\partial \, \varphi} \, ,
\ee
measuring the coupling of $\varphi$ to a particle of type $A$. [For consistency with previous 
work, we keep the notation $\alpha_A$ but warn the reader that this should not be confused 
with the various gauge coupling constants, often denoted $\alpha_i = g_i^2 / 4\pi$.] The 
quantity $\alpha_A$ determines the effect of cosmological matter on the evolution of $\varphi$ 
through the general equation \cite{DN93,DP94}

\be
\label{eq3.2}
\frac{2}{3 - \varphi'^2} \, \varphi'' + \left( 1 - \frac{P}{\rho} \right) \varphi' = - 
\sum_A \alpha_A (\varphi) \, \frac{\rho_A - 3 P_A}{\rho} \, ,
\ee
where the primes denote derivatives with respect to $ p = \ln a + {\rm const} $ and where
$\rho = \Sigma_A \, \rho_A$ and $P = \Sigma_A \, P_A$ 
are the total ``material''
energy density and pressure respectively, both obtained
as sums over the various components  
filling the universe \emph{at the exception of} the kinetic energy density and pressure 
of $\varphi$, $\rho_k = (\widetilde{m}_P^2 /2) (d\varphi/dt)^2 = 
(\widetilde{m}_P^2 /2) H^2 {\varphi'}^2 $ and 
$P_k = \rho_k$. Accordingly, the Friedmann equation reads 
\be \label{friedman}
3 H^2\, =\, \frac{2}{\widetilde{m}_P^2} \, \rho_{\rm tot} \, = \, 
 \frac{2 \, \rho}{ \widetilde{m}_P^2} + 
H^2 {\varphi'}^2.
\ee

Note that $\rho$ and $P$ may also account for the potential energy density and 
pressure of the scalar field,  $\rho_V = V(\varphi)$, $ P_V = - \rho_V $ and that one
can formally extend Eq. (\ref{eq3.2}) to the ``vacuum
energy'' component $V(\varphi)$  by associating to the potential 
$V(\varphi)$ the mass scale $m_V (\varphi) \equiv V(\varphi)^{\frac{1}{4}}$ which gives
$\alpha_V = \partial \ln m_V (\varphi) / \partial \varphi = \frac{1}{4} \, \partial 
\ln V(\varphi) / \partial \, \varphi$. Equation (\ref{eq2.8}) is then recovered in the limit
where the scalar field is the dominant component.

In the simple cases (which are quite frequent; at least as approximate cases) 
where one ``matter'' component, with known ``equation of state'' 
$P_A / \rho_A = w_A = {\rm const.} $,
dominates the cosmological density and pressure, Eq. (\ref{eq3.2}) yields an autonomous
equation for the evolution (with redshift) of $\varphi$. 
Using (\ref{friedman}) one  finds  
that the ``equation of 
state parameter'' $w_{\rm tot} \equiv P_{\rm tot}/ \rho_{\rm tot}$  corresponding to
the {\it total} energy and pressure (including now the kinetic contributions of $\varphi$;
i.e. $\rho_{\rm tot} = \rho + (\widetilde{m}_P^2/2) (d\varphi/dt)^2 \, , \, 
P_{\rm tot} = P + (\widetilde{m}_P^2/2) \, (d\varphi/dt)^2$) is given 
in terms of the ``matter'' equation-of-state parameter $w \equiv P/\rho$ by
\be
\label{eq3.2bis}
w_{\rm tot} = w  + \frac{1 - w}{3} (\varphi')^2 \, . 
\ee
The knowledge of $w_{\rm tot}$ then allows one to write explicitly the energy-balance equation
$ d \rho_{\rm tot} + 3 ( \rho_{\rm tot} + P_{\rm tot}) d \ln a = 0$, which is easily
solved in the simple cases where $w_{\rm tot}$ is (approximately) constant.

We see from Eq.~(\ref{eq3.2}) that, during the radiation era (starting, say, immediately 
after the end of inflation), i.e. when the universe is dominated by an ultra-relativistic gas 
($\rho_A - 3 P_A = 0$), the ``driving force'' on the right-hand side of (\ref{eq3.2}) 
vanishes, so that $\varphi$ is not driven further away towards infinity. Actually, one should 
take into account both the ``inertial'' effect of the ``velocity'' $\varphi'$ acquired during 
the preceding inflationary driving of $\varphi$, and the integrated effect of the many ``mass 
thresholds'', $T_A \sim m_A$, when some component becomes non-relativistic (so that $\rho_A - 
3 P_A \ne 0$). Using the results of \cite{DN93,DP94} one sees that, in our case, both these 
effects have only a small impact on the value of $\varphi$. Therefore, to a good approximation 
$\varphi \simeq \varphi_{\rm end}$ until the end of radiation era.

On the other hand, when the universe gets dominated by non-relativistic  matter, one 
gets a non-zero driving force in Eq.~(\ref{eq3.2}). In the slow roll approximation,
as the transient behaviour has died out, since $w=P/\rho$ gets negligible, we have simply
\be
\label{eq3.3'}
\varphi'_m = - \alpha_m (\varphi) \, ,
\ee
where $\varphi'_m$ stands the $\varphi$-velocity during matter domination.
and $\alpha_m 
 (\varphi)$ denotes the coupling (\ref{eq3.1}) to dark matter.

The coupling to dark matter,  $\alpha_m (\varphi)$, depends on the assumption one makes
about the asymptotic behaviour, at strong bare string coupling,  of the mass of the
WIMPs constituting the dark matter.
One natural looking, minimal assumption is that dark matter, like all visible types of 
matter, is coupled in 
a way which levels off at strong-bare-coupling, as in Eq.~(\ref{eq1.3}). In other words, one 
generally expects that $m_m (\varphi) \simeq m_m (+\infty) (1+b_m \, e^{-c \varphi})$ so that 
$\alpha_m (\varphi) \simeq -b_m \, c \, e^{-c \varphi}$. It is then easy to solve 
Eq.~(\ref{eq3.3'}), with initial conditions $\varphi_0 = \varphi_{\rm end}$, $\varphi'_0 = 0$ 
(inherited from radiation era) at the beginning of the matter era. But we shall not bother to 
write the explicit solution because it is easily seen that the smallness of $e^{-c 
\varphi_{\rm end}}$ guarantees that the ``driving force'' $\propto \alpha_m (\varphi)$ remains 
always so small that the ${\cal O} (10)$ $e$-folds of matter era until vacuum-energy 
domination (or until the present) have only a fractionally negligible effect on $\varphi$.

A more significant evolution of $\varphi$ during the matter era is provided
if, as first proposed in \cite{DGG} and taken up in \cite{SBM,OP}, 
dark matter couples much more strongly to 
$\varphi$ than ``ordinary'' matter. 
Such a stronger coupling to dark matter, which is not constrained by usual equivalence 
principle experiments, follows assuming more general quantum corrections in the 
dark matter sector of the theory, i.e. corrections such that the dark matter mass 
$m_m (\varphi)$, instead of
levelling off, either vanishes or keeps increasing at strong bare coupling:
$m_m(\varphi) \propto e^{c_m \varphi}$, so that $\alpha_m = c_m$ is a (negative or 
positive) constant.
In \cite{GPV} (but see also \cite{AT}) it has been shown that under the latter
assumption (i.e. with a positive coupling parameter $\alpha_m >0 $) 
the dilaton can play the role of quintessence, leading to a late-time cosmology
of accelerated expansion. By Eq. (\ref{eq3.3'}) we have
$\varphi = \varphi_{\rm end} - \alpha_m p $, where $p$ is now counted from the end of 
the radiation era. 
Given that about nine e-folds separate us from the end of the radiation era,
we see that such an evolution might (if  $|\alpha_m|$ is really of order unity)
 have a significant effect on the present value
of $\varphi$ (when compared with the value at the end of inflation, i.e.
$c \varphi_{\rm end} \sim \ln (1/\delta_H) \sim 10$). However, the running of 
$ \varphi$ during the matter era changes the standard recent cosmological picture
and is therefore constrained by observations.
In fact, by equation (\ref{eq3.2bis}), the total matter-era equation of state parameter $w_{\rm tot}$ 
in the presence of the dilaton reads $ w_{\rm tot} = (\varphi'_m)^2/3$. Accordingly, 
the matter density varies as $ \rho \propto a^{- 3 (1 + w_{\rm tot})} = 
a^{-( 3 +(\varphi'_m)^2 )}$, possibly affecting the
standard scenario of structure formation as well as the  global temporal
picture between now and the epoch of matter-radiation equality. 
The compatibility with phenomenology therefore puts constraints on the magnitude of
${\varphi'_m}^2 = \alpha_m (\varphi)^2$.
 In  \cite{GPV}   $w_{\rm tot} < 0.1$, i.e.
$v \equiv {\varphi'_m}^2/ 0.3 < 1$,  was suggested to be  the maximal deviation one can 
roughly tolerate during the matter era, the establishment of a more precise bound being presently
under study \cite{AGUT}.
 We shall
therefore assume either that we are in the ``normal'' case  where the dilaton does not
couple more strongly to dark matter than to ordinary matter (so that 
$\alpha_m (\varphi) \simeq -b_m \, c \, e^{-c \varphi} \ll 1$), or that, if
it does, $ \alpha_m ^2 < 0.3$.
This leads to a displacement of $ \varphi$ during matter era 
smaller than the dispersion  $\varphi_{\rm end} - \varphi_{\rm cl, end} \sim 
\varphi_{\rm cl, end}$ produced by quantum fluctuations during inflation, Eq.~(\ref{eq2.23'}).

In this context one should also
consider the attraction effect of a negative pressure component, either in the form of
a $\varphi$-dependent vacuum energy (dilatonic quintessence) 
or in the form of any other, $\varphi$-independent  component (such as a ``genuine'' 
cosmological constant). Of course, the present recent 
($z \lesssim 1$) accelerated expansion phase is very short (in ``$p$-time'') 
and sensible changes of the dilaton 
value since the end of matter domination are not expected. Still, it is crucial to  
estimate the present dilaton velocity $\varphi'_0$ 
since it is related to the cosmological variations of the coupling 
constants (see next section). 
In the general case where both non-relativistic matter and (possibly 
$\varphi$-dependent) vacuum energy density $V(\varphi)$ are present, the value
of $\varphi_0'$ predicted by our model is obtained by applying Eq. (\ref{eq3.2})
(in the slow-roll approximation):
\be \label{eq2.32bis}
(\Omega_m + \Omega_V)(1-w_0) \varphi'_0 = (\Omega_m + 2 \Omega_V ) \varphi_0' 
= - \Omega_m \alpha_m - 4 \Omega_V \alpha_V.
\ee 

In the above expression $\Omega_m $ and $\Omega_V$ are, respectively, 
the non-relativistic (dark) matter- and the 
vacuum-fraction of critical energy density ($\rho_c \equiv (3/2) {\widetilde{m}_P^2} H^2$),
 and  the already mentioned prescriptions   
$\alpha_V =  \frac{1}{4} \, \partial 
\ln V(\varphi) / \partial \, \varphi$, $P_V = -\rho_V = -V(\varphi)$ have been used.

The value of ${\varphi'_0}$ is therefore some combination of the values of $\alpha_m$ and
$\alpha_V$. We can have two classes of contrasting situations: In the first class,
 the dilaton couples
``normally'' (i.e. weakly) both to dark matter and to dark energy, i.e. both 
 $\alpha_m \simeq -b_m \, c \, e^{-c \varphi} \ll 1$ and $\alpha_V \ll 1$ and 
 Eq.~(\ref{eq2.32bis}) implies $\varphi'_0 \ll 1$. In the second class, the dilaton
 couples more strongly to some type of dark matter or energy, i.e. either (or both) 
 $\alpha_m$  or/and   $\alpha_V$ is of order unity so that ${\varphi'_0} = {\cal O} (1)$.
The second case is realized in the scenario of \cite{GPV}. In the context of this 
scenario we have an exponential dependence of the potential on $\varphi$,
$ V(\varphi) \simeq V_1 e^{-c\varphi}$ so that $\alpha_V \simeq - (c/4)$ and
\be \label{eq2.34}
\varphi'_0 \, = \, \frac{c\, \Omega_V - \alpha_m \Omega_m}{2\Omega_V + \Omega_m} \, \lesssim  \,
\frac{c}{3}.
\ee 
The last inequality follows from the bound
$\alpha_m >c/2$ (which is a necessary condition to have positive
acceleration in the model \cite{GPV}) and the reasonable bound $\Omega_m > 0.25$. 
 
  In the present work, we
 wish, however, to be as independent as possible from specific assumptions (as the ones
 used in \cite{GPV}). Therefore, rather than insisting on specific (model-dependent) predictions
 for the present value of ${\varphi'_0}$ we wish to find the (model-independent) upper
bounds on the possible values of ${\varphi'_0}$ set by current observational data. 
There are several ways of getting such phenomenological bounds, because the existence
of a kinetic energy (and pressure) associated to $d \varphi / d t = H \varphi'$ has
several observable consequences. A rather secure bound can be obtained by relating
the value of $\varphi'$ to the  deceleration parameter 
$q \equiv - \ddot{a}a/\dot{a}^2$. In the general class of models that we consider,
 the cosmological
energy density and pressure have (currently) three significant contributions: dark matter
( $\Omega_m = \rho_m/ \rho_c$), dark energy ($\Omega_V$) and the kinetic effect of a
scalar field ($\Omega_k = \rho_k/\rho_c$ with $\rho_k = (\widetilde{m}_P^2 /2) (d\varphi/dt)^2 = 
(\widetilde{m}_P^2 /2) H^2 {\varphi'}^2 $ so that $\Omega_k = \varphi'^2/3$).
We assume (consistently with recent cosmic background data) that the space curvature is
zero. Therefore we have the first relation 
\be \label{om1}
 \Omega_m + \Omega_V + \Omega_k = 1 = \Omega_m + \Omega_V + \varphi'^2/3 \, .
\ee 
The deceleration parameter is given by the general expression
$ 2 q = \Sigma_A \Omega_A ( 1 + 3 w_A)$. Using $ w_m = 0$, $ w_V = -1$ and $w_k = + 1$,
we get 
\be \label{q1}
2 q  = \Omega_m - 2 \Omega_V + \frac{4}{3} \varphi'^2.
\ee
Using the relation (\ref{om1})  above to eliminate $\Omega_V$ we get the following expression for 
$\varphi'^2$ in terms of the observable quantities $q$ and $\Omega_m$
\be \label{q2}
\varphi'^2 = 1 + q - \frac{3}{2} \Omega_m.
\ee

The supernovae Ia data \cite{SNI} give a strict upper bound on the
present value $q_0$: $q_0<0$. A generous lower bound on the present value of
$\Omega_m$ is $\Omega_{m 0} > 0.2$ \cite{omegam}. 
Inserting these two constraints in Eq.(\ref{q2}) finally yields
the safe upper bound
\be \label{q3}
 {\varphi'_0}^2 < 0.7 \, , \; {\rm i.e.} \; \vert \varphi'_0 \vert < 0.84 \; .
\ee
 
 To summarize, 
quite different rates of evolution for the dilaton are possible. A very slow variation
is expected whenever dilaton couplings to both dark energy and dark matter follow
the ``normal'' behaviour (\ref{eq1.3}). Otherwise, dilaton variations on the Hubble 
scale are expected. However, cosmological observations set the strict
upper bound (\ref{q3}) on the present time variation of $\varphi$.
For the purpose of the present section (evaluating the current location of the dilaton)
these two alternatives do not make much difference because the vacuum-dominance
era has started less than about 0.7 e-folds away ($\ln (1+z_*)$ with $z_* < 1$).
Therefore, $\varphi$ did not have enough ``$p$-time'' , during vacuum dominance, to
move much, even if it is coupled to vacuum energy with $\alpha_V \simeq - (c/4) \sim 1$.

 Finally, we conclude from this analysis that, to a 
good approximation (and using the fact that the phenomenology of
the matter-era constrains the dark-matter couplings of the dilaton to be rather small),
 the value of $\varphi$ now is essentially given by the value $\varphi_{\rm 
end}$ at the end of inflation, i.e. by Eq.~(\ref{eq2.23}).

\section{Deviations from general relativity induced by a runaway dilaton}\label{sec3}

\subsection{Composition-independent deviations from general relativity}

The previous section has reached the conclusion that present deviations 
from general relativity are given, to a 
good approximation, by the values of the matter-coupling coefficients 
$\alpha_{A} (\varphi)$ 
given by Eq.~(\ref{eq3.1}) calculated at $\varphi \simeq \varphi_{\rm end}$ as given by 
Eq. (\ref{eq2.23'}). Let us now see the 
meaning of this result in terms of observable quantities.

Let us first consider the (approximately) composition-independent deviations from general 
relativity, i.e. those that do not essentially depend on violations of the equivalence 
principle. Most composition-independent gravitational experiments (in the solar system or in 
binary pulsars) consider the long-range interaction between objects whose masses are 
essentially baryonic (the Sun, planets, neutron stars). As argued in \cite{TV88,DP94} the 
relevant coupling coefficient $\alpha_A$ is then approximately universal and given by the 
logarithmic derivative of the QCD confinement scale $\Lambda_{\rm QCD} (\varphi)$, because the 
mass of hadrons is essentially given by a pure number times $\Lambda_{\rm QCD} (\varphi)$. [We 
shall consider below the small, non-universal, corrections to $m_A (\varphi)$ and 
$\alpha_A (\varphi)$ 
linked to QED effects and quark masses.] Remembering from Eq.~(\ref{eq1.1}) the 
fact that, in the string frame (where there is a fixed cut-off linked to the string mass 
$\widetilde{M}_s \sim (\alpha')^{-1/2}$) the gauge coupling is dilaton-dependent ($g_F^{-2} = 
B_F (\varphi)$), we see that (after conformal transformation) the Einstein-frame confinement 
scale has a dilaton-dependence of the form
\be
\label{eq3.5}
\Lambda_{\rm QCD} (\varphi) \sim C^{-1/2} \, B_g^{-1/2} (\varphi) \exp [- 8 \pi^2 \, b_3^{-1} 
\, B_F (\varphi)] \, \widetilde{M}_s \, ,
\ee
where $b_3$ denotes the one-loop (rational) coefficient entering the Renormalization Group 
running of $g_F$. Here $B_F (\varphi)$ denotes the coupling to the ${\rm SU} (3)$ gauge 
fields. For simplicity, we shall assume that (modulo rational coefficients) all gauge fields 
couple (near the string cut off) to the same $B_F (\varphi)$. This yields the following 
approximately universal dilaton coupling to hadronic matter
\be
\label{eq3.6}
\alpha_{\rm had} (\varphi) \simeq \left( \ln \left( \frac{\widetilde{M}_s}{\Lambda_{\rm QCD}} 
\right) + \frac{1}{2} \right) \frac{\partial \ln B_F^{-1} (\varphi)}{\partial \, \varphi} \, .
\ee
We recall that the quantity $\alpha_{\rm had}(\varphi)$, 
which measures the coupling of the
dilaton to hadronic matter, should not be confused with any "strong" gauge
coupling, $\alpha_s = g_s^2/4 \pi$.
Numerically, the coefficient in front of the R.H.S. of (\ref{eq3.6}) is of order 40. 
Consistently with our basic assumption (\ref{eq1.3}), we parametrize 
the $\varphi$ dependence of the gauge coupling $g_F^2 = B_F^{-1}$ as
\be
\label{eq3.7}
B_F^{-1} (\varphi) = B_F^{-1} (+ \infty) \, [1 - b_F \, e^{-c\varphi}] \, .
\ee
Note that, like $b_\lambda$ (see section \ref{sec2}), 
also $b_F$ is expected to be smallish [$\sim B_F^{-1} (+ \infty)$ or, equivalently,
$\sim C_F^{-1}$ in the notations of (\ref{eq1.3})] and typically the ratio 
$b_F/b_\lambda$ is of order unity.
We finally obtain 
\be
\label{eq3.8}
\alpha_{\rm had} (\varphi) \simeq 40 \, b_F \, c \, e^{-c\varphi} \, .
\ee
We can now insert the estimate (\ref{eq2.23'}) of the value of $\varphi$ reached because of the 
cosmological evolution. Neglecting the ${\cal O} (1)$ renormalization factor due to quantum noise,
we get the estimate
\be
\label{eq3.9}
\alpha_{\rm had} (\varphi_{\rm end}) \simeq 3.2 \, \frac{b_F}{b_{\lambda} \, c} \, 
\delta_H^{\frac{4}{n+2}} \, ,
\ee
\be
\label{eq3.10}
\alpha_{\rm had}^2 (\varphi_{\rm end}) \simeq 10 \left(\frac{b_F}{b_{\lambda} \, c} \right)^2 
\delta_H^{\frac{8}{n+2}} \, .
\ee
As said above, it is plausible to expect that the quantity $c$ (which is a ratio) and the 
ratio $b_F / b_{\lambda}$ are both of order unity. This then leads to the numerical estimate 
$\alpha_{\rm had}^2 \sim 10 \, \delta_H^{\frac{8}{n+2}}$, with $\delta_H \simeq 5 \times 
10^{-5}$. An interesting aspect of this result is that the expected present value of 
$\alpha_{\rm had}^2$ depends rather strongly on the value of the exponent $n$ (which entered 
the inflaton potential $V(\chi) \propto \chi^n$). In the case $n=2$ (i.e. $V(\chi) = 
\frac{1}{2} \, m_{\chi}^2 \, \chi^2$) we have $\alpha_{\rm had}^2 \sim 2.5 \times 10^{-8}$, 
while if $n=4$ ($V (\chi) = \frac{1}{4} \, \lambda \, \chi^4$) we have $\alpha_{\rm had}^2 \sim 
1.8 \times 10^{-5}$.

 How do these numbers compare to present (composition-independent) 
experimental limits on deviations from Einstein's theory \cite{exp}? This question has been 
addressed in the literature. Concerning solar-system (post-Newtonian) tests it was shown (see, 
e.g., \cite{DEF92}) that the two main ``Eddington'' parameters $\gamma - 1$ and $\beta - 1$ 
measuring post-Newtonian deviations from general relativity are linked as follows to the 
dilaton coupling $\alpha_{\rm had} (\varphi)$:
\be
\label{eq3.11}
\gamma - 1 = - 2 \, \frac{\alpha_{\rm had}^2}{1 + \alpha_{\rm had}^2} \simeq - 2 \, \alpha_{\rm 
had}^2 \, ,
\ee
\be
\label{eq3.12}
\beta - 1 = \frac{1}{2} \, \frac{\alpha'_{\rm had} \, \alpha_{\rm had}^2}{(1 + \alpha_{\rm 
had}^2)^2} \simeq \frac{1}{2} \, \alpha'_{\rm had} \, \alpha_{\rm had}^2 \, ,
\ee
where $\alpha'_{\rm had} \equiv \partial  \alpha_{\rm had} (\varphi) / \partial \varphi$.

{}From Eq.~(\ref{eq3.8}) we see that $\alpha'_{\rm had} \simeq -c \, \alpha_{\rm had}$, so that 
the deviation $\beta - 1$ is ${\cal O} (\alpha_{\rm had}^3)$ and thereby predicted to be too 
small to be phenomenologically interesting. This leaves $\gamma - 1 \simeq -2 \, \alpha_{\rm 
had}^2$ as the leading observable deviation. The best current solar-system limit on $\gamma - 
1$ comes from Very Long Baseline Interferometry measurements of the deflection of radio waves 
by the Sun and is (approximately) $\vert \gamma - 1 \vert \lesssim 2 \times 10^{-4}$, 
corresponding to $\alpha_{\rm had}^2 \lesssim 10^{-4}$ (see \cite{exp} for reviews and 
references). In addition to solar-system tests, we should also consider binary-pulsar tests 
which provide another high-precision window on possible deviations from general relativity. 
They have been analyzed in terms of the two quantities $\alpha_{\rm had}$ (denoted $\alpha$) 
and $\alpha'_{\rm had}$ (denoted $\beta$) in \cite{DEF}. The final conclusion is that the 
binary-pulsar limit on $\alpha_{\rm had}$ is of order $\alpha_{\rm had}^2 \lesssim 10^{-3}$.

At this stage it seems that the runaway scenario explored here is 
leading to deviations from 
general relativity which are much smaller than present experimental limits. However, we must 
turn our attention to {\it composition-dependent} effects which turn out to be much more 
sensitive tests.

\subsection{Composition-dependent deviations from general relativity}

Let us then consider situations where the non-universal couplings of the dilaton induce 
(apparent) violations of the equivalence principle. Let us start by considering the 
composition-dependence of the dilaton coupling $\alpha_A$, Eq.~(\ref{eq3.1}), i.e. the 
dependence of $\alpha_A$ on the type of matter we consider.
The definition of $\alpha_A$ is such that, at the Newtonian approximation, the interaction 
potential between particle $A$ and particle $B$ is $-G_{AB} \, m_A \, m_B / r_{AB}$ where 
\cite{DP94}
\be
\label{eq3.13}
G_{AB} = G (1 + \alpha_A \, \alpha_B) \, .
\ee
Here, $G$ is the bare gravitational coupling constant entering the Einstein-frame action 
(\ref{eq2.4}), and $\alpha_A = \alpha_A (\varphi)$ is the strength of the dilaton coupling to 
$A$-particles, taken at the present (cosmologically determined) VEV of $\varphi$. The term 
$\alpha_A \, \alpha_B$ comes from the additional attractive effect of dilaton exchange. Two 
test masses, made respectively of $A$- and $B$-type particles will then fall in the 
gravitational field generated by an external mass $m_E$ with accelerations differing by
\be
\label{eq3.14}
\left( \frac{\Delta a}{a} \right)_{AB} \equiv 2 \, \frac{a_A - a_B}{a_A + a_B} \simeq 
(\alpha_A - \alpha_B) \, \alpha_E \, .
\ee
We have seen above that in lowest approximation $\alpha_A \simeq \alpha_{\rm had}$ does not 
depend on the composition of $A$. We need, however, now to retain the small 
composition-dependent effects to $\alpha_A$ linked to the $\varphi$-dependence of QED and 
quark contributions to $m_A$. This has been investigated in \cite{DP94} with the result
\be
\label{eq3.15}
\left( \frac{\Delta a}{a} \right)_{AB} = \left( \frac{\alpha_{\rm had}}{40} \right)^2 \left[ 
C_B \, \Delta \left( \frac{B}{M} \right) + C_D \, \Delta \left( \frac{D}{M} \right) + C_E \, 
\Delta \left( \frac{E}{M} \right) \right]_{AB} \,  ,
\ee
where $(\Delta X)_{AB} \equiv X_A - X_B$, where $B \equiv N + Z$ is the baryon number, $D 
\equiv N-Z$ the neutron excess, $E \equiv Z (Z-1) / (N+Z)^{1/3}$ a quantity linked to nuclear 
Coulomb effects, and where $M \equiv m/u$ denotes the mass in atomic mass unit, $u = 
931.49432$ MeV. It is difficult (and model-dependent) to try to estimate the coefficients 
$C_B$ and $C_D$. It was argued in \cite{DP94} that their contributions to (\ref{eq3.15}) is 
generically expected to be sub-dominant with respect to the last contribution, $\propto C_E$, 
which can be better estimated because it is linked to the $\varphi$-dependence of the 
fine-structure constant $e^2 \propto B_F^{-1} (\varphi)$. This then leads to the numerical 
estimate $C_E \simeq 3.14 \times 10^{-2}$ and a violation of the universality of free fall 
approximately given by
\be
\label{eq3.16}
\left( \frac{\Delta a}{a} \right)_{AB} \simeq 2 \times 10^{-5} \, \alpha_{\rm had}^2 \left[ 
\left( \frac{E}{M} \right)_A - \left( \frac{E}{M} \right)_B \right] \, .
\ee

The values of $B/M$, $D/M$ and $E/M$ have been computed in \cite{D96}. For mass-pairs that 
have been actually used in recent experiments (such as Beryllium and Copper), as well as for 
mass-pairs that are planned to be used in forthcoming experiments (such as Platinum and 
Titanium) one finds: $(E/M)_{\rm Cu} - (E/M)_{\rm Be} = 2.56$, $(E/M)_{\rm Pt} - (E/M)_{\rm 
Ti} = 2.65$. Using the average estimate $\Delta (E/M) \simeq 2.6$, we get from (\ref{eq3.16}) 
and (\ref{eq3.10}) the estimate
\be
\label{eq3.17}
\left( \frac{\Delta a}{a} \right) \simeq 5.2 \times 10^{-5} \, \alpha_{\rm had}^2 \simeq 5.2 
\times 10^{-4} \left( \frac{b_F}{b_{\lambda} \, c} \right)^2 \, \delta_H^{\frac{8}{n+2}} 
\, .
\ee
Note also (from (\ref{eq3.11})) the link between composition-dependent effects and 
post-Newtonian ones
\be
\label{eq3.18}
\left( \frac{\Delta a}{a} \right) \simeq - 2.6 \times 10^{-5} (\gamma - 1) \, .
\ee
As current tests of the universality of free fall (UFF) have put limits in the $10^{-12}$ 
range (e.g. $(\Delta a / a)_{\rm Be \, Cu} = (-1.9 \pm 2.5) \times 10^{-12}$ from 
\cite{Su94}), we see from Eq.~(\ref{eq3.18}) that this corresponds to limits on $\gamma - 1$ 
or $\alpha_{\rm had}^2$ in the $10^{-7}$ range. Therefore tests of the UFF put much more 
stringent limits on dilaton models than solar-system or binary-pulsar tests.

If we insert the estimate $\delta_H \sim 5 \times 10^{-5}$ in (\ref{eq3.17}) we obtain a level 
of violation of UFF due to a runaway dilaton which is
\be
\label{eq3.19}
\frac{\Delta a}{a} \simeq 1.3 \left( \frac{b_F}{b_{\lambda} \, c} \right)^2 \times 
10^{-12} \quad \hbox{for} \ n=2 \, ,
\ee
\be
\label{eq3.20}
\frac{\Delta a}{a} \simeq 0.98 \left( \frac{b_F}{b_{\lambda} \, c} \right)^2 \times 
10^{-9} \quad \hbox{for} \ n=4 \, .
\ee
At face value, one is tempted to conclude that a scenario with $n=4$ (i.e. $V(\chi) \propto 
\chi^4$) tends to be too weak an attractor towards $\varphi = + \infty$ to be naturally 
compatible with equivalence-principle tests. [See, however, the discussion below.] On the 
other hand, the simple scenario $n=2$ ($V(\chi) = \frac{1}{2} \, m_{\chi}^2 \, \chi^2$) is 
quite appealing in that it naturally provides enough attraction towards $\varphi = + \infty$ 
to be compatible with all existing experimental tests. At the same time it suggests 
that a modest improvement in the precision of UFF experiments might discover a violation 
caused by a runaway dilaton.

\subsection{Cosmological variation of ``constants''}

Let us now consider another possible deviation
from General Relativity and the standard model: a possible variation 
of the coupling constants, most notably of the fine structure 
constant $e^2/\hbar c$ on which the strongest limits are available.
We will discuss first the effects due to the cosmological time-variation of the 
homogeneous component of $\varphi$ 
and, in the next subsection, the possible spatial (and time) variations 
due to  quantum fluctuations of $\varphi$
 as they got amplified during inflation.

 Consistently with our previous assumptions we expect 
$e^2 \propto B_F^{-1} (\varphi)$ so that, from (\ref{eq3.7}),
\be
\label{eq3.21}
e^2 (\varphi) = e^2 (+ \infty) \, [1 - b_F \, e^{- c \varphi} ] \, .
\ee
The present logarithmic variation of $e^2$ (using again $dp = H \, dt ;  
\varphi' = d \varphi / d p$) is thus
given by
\be
\label{eq3.21bis}
\frac{d \ln e^2}{H \, dt} = \frac{d \ln e^2}{dp} \simeq b_F \, c \, e^{-c \varphi} \, 
\varphi'_0 \, ,
\ee
where the current value of $\varphi'$,  $\varphi'_0$, is given in general by Eq.~(\ref{eq2.32bis}).
Using  Eq.~(\ref{eq3.8}), we can rewrite the result (\ref{eq3.21bis}) in terms of the 
hadronic coupling:
\be
\label{eq3.21ter}
\frac{d \ln e^2}{H \, dt} \simeq \frac{1}{40} \alpha_{\rm had} \varphi'_0 \, .
\ee

As said in section \ref{sec2b}, we have basically two alternatives concerning the 
current coupling of the dilaton
to the dominant energy sources in the universe. 
These two alternatives lead to drastically different predictions for the 
current value of the rate of variation of the fine-structure constant. 
We shall consider
these two alternatives in turn.

In the conservative case where 
the dilaton does not play any special role in the present accelerated phase of the 
universe ($\alpha_V \simeq 0$) nor does it have any stronger coupling to dark matter 
than to visible matter ($\alpha_m \simeq -b_m \, c \, e^{-c\varphi}$) the dilaton 
``velocity'' $\varphi'$ is exponentially suppressed ( so that, from (\ref{om1}),
$ \Omega_V \simeq  1 - \Omega_m$) and 
by Eq. (\ref{eq2.32bis}) one obtains
\be \label{eq3.22'}
\frac{d \ln e^2}{H \, dt} \, \simeq \,  - \frac{\Omega_m}{\Omega_m + 2 \Omega_V} \, b_F \, c\,  
e^{-c\varphi}\alpha_m(\varphi) \, \simeq \, \frac{\Omega_m}{2-\Omega_m} \, b_F \, b_m \, c^2 \, 
e^{-2 c\varphi}.
\ee
An indicative value for the ratio $\Omega_m /(\Omega_m + 2 \Omega_V) \simeq 
\Omega_m /(2-\Omega_m)$, by taking for instance
$\Omega_m = 0.3$, is $0.18$. 
As above, it is useful to relate (\ref{eq3.22'}) to the estimate (\ref{eq3.8}) for $\alpha_{\rm had}$. 
This yields
\be
\label{eq3.23}
\frac{d \ln e^2}{H \, dt} \, \simeq \, \frac{1}{(40)^2} \, \frac{\Omega_m}{2-\Omega_m} \, \frac{b_m}{b_F} \, 
\alpha_{\rm had}^2 \, .
\ee
In terms of the UFF level $\Delta a / a$ predicted by our model in (\ref{eq3.17}) we see also 
that
\be
\label{eq3.24}
\frac{d \ln e^2}{H \, dt} \, \simeq \, 
12 \, \frac{\Omega_m}{2-\Omega_m} \, \frac{b_m}{b_F} \, \frac{\Delta a}{a} \, .
\ee
Even  if the universe were completely dominated by dark matter 
($\Omega_m=1$) we see, assuming that $b_m/b_F$ is of order unity,
 that current experimental limits on UFF 
($\Delta a / a \lesssim 
10^{-12}$) imply (within dilaton models) that $\vert d \ln e^2 / dt \vert \lesssim 10^{-11} \, 
H \sim 10^{-21} \, {\rm yr}^{-1}$ (the sign of $d \ln e^2 / dt$ being given by the
sign of $b_m/b_F$). This level of variation is much smaller than the current best 
limit on the time variation of $e^2$, namely $\vert d \ln e^2 / dt \vert \lesssim 5 \times 
10^{-17} \, {\rm yr}^{-1} \sim 5 \times 10^{-7} \, H$, as obtained from an analysis of Oklo data 
\cite{Oklo}. (Note that the assumption-dependent analysis of Ref. \cite{OPQCCV}
gives a limit on the variation of $e^2$ which is strengthened by 
about two orders of magnitude.)

The situation, however, is drastically different if we consider the alternative case
where the dilaton coupling to the current dominant energy sources does not tend to 
triviality, as in the case of a $\varphi$-dependent vacuum energy
$V(\varphi) = V_0 + V_1 e^{-c\varphi}$ when the first 
term is zero or negligible.
In such a case the dilaton shares a relevant part of the 
total energy density and more significant (though still quite
constrained by UFF data) variations of 
the coupling constants are generally expected.
A general expression for the dilaton ``velocity'' is given in eq.
(\ref{q2}) in terms of observable quantities.  
Using Eqs.~(\ref{q2}) and (\ref{eq3.21ter}) one can relate the expected variation 
of the electromagnetic coupling constant to the hadronic coupling:
\be \label{eq3.25'}
\frac{d \ln e^2}{H \, dt} \simeq \pm \, \frac{\alpha_{\rm had}}{40}  
\, \sqrt{  1 + q_0- 3\Omega_m/2  } 
 \, .
\ee 

We can also use the estimate (\ref{eq3.9}) relating
${\alpha}_{\rm had}$ to the density fluctuations generated during inflation.
We obtain
\be
\label{eq3.26}
\frac{d \ln e^2}{H \, dt} \simeq  
\pm \, 8 \times 10^{-2}  \, \sqrt{  1 + q_0- 3\Omega_m/2  }
\, \frac{b_F}{b_{\lambda} c} \,  \delta_H^{\frac{4}{n+2}}  \, .
\ee 

However, in view of the theoretical uncertainties attached to the initial 
conditions
$\chi_{\rm in}$ and $\varphi_{\rm in}$ used in the estimate (\ref{eq3.9}), 
as well as  the ones associated to the order unity ratio 
$b_F / (b_{\lambda} c)$,
it is more interesting
to rewrite our prediction in terms of {\it observable} quantities. Using again
the link Eq.~(\ref{eq3.17}) between ${\alpha}_{\rm had}$ and the observable
violation of the universality of free fall (UFF) the 
above result can be written in the form
\be
\label{eq3.26''}
\frac{d \ln e^2}{H \, dt} \simeq \pm \, 3.5 \times 10^{-6} \, 
\sqrt{  1 + q_0- 3\Omega_m/2  }
\, \sqrt{ 10^{12} \frac{\Delta a}{a}} \, .
\ee

Note that the sign of the variation of $e^2$ is in general model-dependent (as it
depends both on the sign of $b_F$ and the sign of $\varphi'_0$). Specific classes
of models might, however, favour particular signs of $d e^2/d t$. For instance,  
 from the point of view of \cite{V01} one would expect the ${\cal O} (e^{-\phi})$ terms
 in Eq.~(\ref{eq1.3}) to be positive, which would then imply that $b_F$ is positive.
 If we combine this information with the prediction Eq.~(\ref{eq2.34}) of the
model \cite{GPV} implying that $\varphi'$ is also positive, we would reach the conclusion that 
$e^2$ must be currently {\it increasing}. 

Independently of this question of the sign, we see that
Eq.~(\ref{eq3.26''}) predicts an interesting link between the observational violation 
of the UFF
(constrained to $\Delta a/a \lesssim 10^{-12}$),
and the current time-variation of the fine-structure constant. Contrary to the relation 
(\ref{eq3.24}), obtained above under the alternative assumption about the dilaton dependence
of the dominant cosmological energy, which predicted a  relation linear in $\Delta a/a$,
 we have here a relation involving the square root of the UFF violation (such
a relation is similar to the result of \cite{DGG} which concerned the time-variation of
the Newton constant). 

The phenomenologically interesting consequence of Eq.~(\ref{eq3.26''})
is to predict a time-variation of constants which may be large enough to be detected
by high-precision laboratory experiments. Indeed, using $H_0 \simeq 66$ km/s/Mpc, and the
plausible estimates
$\Omega_m = 0.3$, $q_0 = -0.4$,
Eq.~(\ref{eq3.26''}) yields the numerical estimate 
$d \ln e^2 / dt \sim \pm 0.9 \times 10^{-16}
\,   \sqrt{ 10^{12} \Delta a/a} \, {\rm yr}^{-1}$.
 Therefore, the current bound on UFF violations
($\Delta a/a \sim 10^{-12}$) corresponds to the level $10^{-16}{\rm yr}^{-1}$,
which is comparable to the planned sensitivity of currently developed cold-atom clocks
\cite{salomon}. [Present laboratory bounds are at the  $10^{-14}{\rm yr}^{-1}$
level \cite{prestage,salomon}.] Note that if we insert in Eq.~(\ref{eq3.26''}) 
the secure bounds $\Omega_m > 0.2$ and $q_0 < 0$ (leading to the limit Eq.~(\ref{q3})), 
we get as maximal estimate of the time
variation of the fine-structure constant $d \ln e^2 / dt \sim \pm 2.0 \times 10^{-16}
\,   \sqrt{ 10^{12} \Delta a/a} \, {\rm yr}^{-1}$.
We note also that the upper limit on the variation of $e^2$ 
given by the Oklo data, i.e. 
$\vert d \ln e^2 / dt \vert \lesssim 5 \times 10^{-17} \, {\rm yr}^{-1}$ \cite{Oklo},
``corresponds''  to a violation of the UFF at the
level  $ \sim  10^{-13}$. 

In this respect, it is interesting to consider not only the {\it present} variation of
$e^2$ (the only one relevant for laboratory experiments), but also its variation
over several billions of years. (We recall that the  Oklo phenomenon took place 
about two billion years ago, and that astronomical observations constrain the variation
of $e^2$ over the last ten billion years or so). In particular, an interesting question
is to see whether our model could reconcile the Oklo limit (which corresponds to a 
redshift $z \simeq 0.14$) with the recent claim \cite{Webb}
of a variation $ \Delta e^2/ e^2 = ( -0.72 \pm 0.18 ) \times 10^{-5}$ around
redshifts $ z \approx 0.5-3.5$ as proposed in \cite{SBM,OP}. 
The only hope of reconciling the two results would be to
allow for a faster variation of $e^2$ for redshifts $ z > 0.5$. Such recent redshifts have
(apparently) been connected to a transition from matter dominance to vacuum dominance. Let us
see whether taking into account this transition might allow for a large enough change of
$e^2$ around redshifts $ z \approx 0.5-3.5$. We must clearly assume the ``strong coupling'' scenario
$ \alpha_m = {\cal O}(1)$. In this scenario, the variation of $\varphi$
during the matter era is given by Eq.(\ref{eq3.3'}). Neglecting, for simplicity, the transient 
evolution effects localized around the matter-vacuum transition (and treating  both
$\varphi'_m = - \alpha_m$ and $\varphi'_V = \varphi'_0$ as constants), the solution giving the 
recent cosmological evolution of $ \varphi $ reads
$ \varphi - \varphi_0 = - \varphi'_0 \ln (1+z)$ during the vacuum era, and
$ \varphi - \varphi_0 = - \varphi'_0 \ln (1+z_*) - \varphi'_m \ln [ (1+z)/(1+z_*)]$ during
the matter era (the index $0$ refers to the present epoch, i.e. $z=0$;  $z_*$ denotes 
the transition redshift).
 Inserting this change in Eq.(\ref{eq3.21}) leads to the following expression 
for the cosmological change of the fine-structure constant:
\be
\label{3.26''}
\frac{e^2 - e^2_0}{e^2_0} = - {\rm sign} (b_F) \, 3.5 \times 10^{-6} \,
\lbrack \varphi'_0 \ln (1+z_*) + \varphi'_m \ln \frac{ 1+z}{1+z_*} \rbrack \,  
\sqrt{ 10^{12} \frac{\Delta a}{a}}\, .
\ee
Here, we have written the result for the matter era. During the vacuum era the bracket is
simply $[ \varphi'_0 \ln (1+z) ]$. Remembering that the absolute value of $\varphi'_m$ is 
(like that of $\varphi'_V$)
observationally constrained to be smaller than $\sqrt{0.3} \simeq 0.55$ (and that $\varphi'_0$
is also constrained by $ \vert \varphi'_V \vert < 0.84$), we see that there is
no way, within our model, to explain a variation of $e^2$ as large as 
$ \Delta e^2/ e^2 = ( -0.72 \pm 0.18 ) \times 10^{-5}$ around redshifts $ z \approx 0.5-3.5$ 
\cite{Webb}. In our model, even under the assumption that UFF is violated just below
the currently tested level, such a change would have to correspond to a 
value  $ \vert \varphi'_m \vert > 2$, entailing observationally unacceptable modifications
of standard cosmology. [For instance, in the model \cite{GPV} a value as large as
 $\alpha_m >1$ already leads to a pathological behaviour (``total dragging'') 
where all the components scale like radiation.] This difficulty of reconciling 
the Oklo limit with the claim of 
\cite{Webb} was addressed in \cite{OP,SBM} within a different class of models, namely
with a field $\phi$ 
which {\it does not} couple universally to all gauge fields $F_{\mu \nu}$, as the
dilaton $\varphi$ is expected to do. The fact that the field $\phi$ in \cite{OP} 
(or $\psi$ in \cite{SBM}) is assumed to couple only to the electromagnetic 
gauge field
drastically changes our Eq. (\ref{eq3.17}) and allows one to satisfy the UFF limit
$\Delta a/a \lesssim 10^{-12}$, for a stronger coupling of
$\phi$ to electromagnetism than in our class of models, i.e. (in our notation)
for a larger  $d \ln B_F(\varphi)/ d \varphi$. This explains why Ref. \cite{OP} 
could construct some explicit (but fine-tuned) models in which all
observational
limits (UFF, Oklo,...) could be met and still allow for a variation of
$e^2$ as strong as the claim[23].
The maximal variation predicted by Eq.(\ref{3.26''})
for redshifts  corresponding to the matter era
(obtained when $\Delta a/a = 10^{-12}$ and  $\varphi'_m = \pm \sqrt{0.3}$; and assuming a smaller
value of $\varphi'_0$ to be compatible with the Oklo constraint), is of order
$ \Delta e^2/ e^2 = \pm 1.9 \times 10^{-6}$. This is only a factor $\sim 4$ below the claim \cite{Webb}
and is at the level of their one sigma error bar. Therefore a modest improvement in the
observational precision (accompanied by an improved control of systematics) will start to
probe a domain of variation of constants which, according to our scenario, corresponds
to an UFF violation smaller than the $10^{-12}$ level.

\subsection{Spatio-temporal fluctuations of the ``constants''}

We now turn to the second possible source of spatial-temporal variations 
for $e^2$ in our model,
the quantum fluctuations of the dilaton generated during inflation.  
Within linear perturbation theory, the relevant calculation may
be summarized  as follows.

 Consider a flat FRW universe 
$ds^2 = -dt^2 + a(t)^2 \, \Sigma_i dx_i^2 $ . 
The dilaton fluctuations 
can be expanded in Fourier components $\delta \varphi_{\bf k}$ of given comoving
momentum ${\bf k}$ as follows:
\be
\delta\varphi({\bf x},t) = \frac{1}{(2\pi)^{3/2}}\int d^3k \,
\delta \varphi_{\bf k}(t) e^{i {\bf k x}},
\ee
where $t$ is the cosmological time.  
Each Fourier mode $\delta \varphi_{\bf k}$ 
``leaves'' the horizon during inflation with an amplitude 
$\sim \whh_{\rm ex}(k)/\sqrt{2 k^3}\, $ 
\cite{lyth} where, by definition, $\whh_{\rm ex}(k)$ is the value of the dimensionless
Hubble expansion rate as $k a^{-1}$ equals $H$ during inflation (note that we denote here
$\whh_{\rm ex}$ what was denoted $\whh_{\times}$ above).
Well after the exit ($k \ll a H$) the amplitude of each mode ``freezes out'',
i.e. remains roughly constant, until it reenters the horizon during the 
post-inflationary epoch 
($k a_{re}^{-1} \simeq H_{re}$). After re-entry  the amplitude starts to 
damp out as $a^{-1}$. 
For a given Fourier mode $\delta\varphi_{\bf k}(t)$, 
the latter damping effect is described by the piecewise function 
\begin{equation}
f_z(k) \equiv \left\{ \begin{array}{llc}
1 & {\rm if} & \quad a_0^{-1} H_0^{-1} \, k \, < \, (z+1)^{1/2} \\[2mm]
a_0^2 H_0^2 \, (z+1) k^{-2}& {\rm if} & \qquad (z+1)^{1/2} \, < \, 
a_0^{-1} H_0^{-1}\, k \, < \, 10^2  \\[2mm]
10^{-2} a_0 H_0 \, (z+1) k^{-1}\qquad \qquad & {\rm if} &  \quad a_0^{-1} H_0^{-1}\, k \, 
> \, 10^2 
\end{array} \right.
\ee
Here the cosmological redshift $z=a_0/a(t)-1$ has been introduced in replacement of the 
cosmological time
$t$. The first case refers to  Fourier modes that have
 not reentered yet at redshift $z$ 
and whose amplitudes are still frozen. The second and third cases refer to  modes 
that reenter during matter and radiation domination respectively.
Putting all together, and assuming a gaussian probability distribution for the 
perturbations, we have:
\be \label{eq3.28f}
\langle\delta\varphi_{\bf k} (t)^* \, \delta\varphi_{\bf k'} (t') \rangle  \, \, =\, 
\frac{\whh_{\rm ex}^2(k)}{2k^3} \, f_z(k)\, f_{z'}(k)
\, \delta^3({\bf k} - {\bf k'}) \, .
\ee
Possible spatial/temporal variations of $e^2$ induced by the 
fluctuations of the dilaton will be given by
\be \label{eq3.27}
\left. \frac{\Delta^{\rm fluc} e^2}{e^2}\right|_{({\bf x}, t;\, {\bf x'}, t')} \, = \, 
\frac{d \ln e^2}{d\varphi}\, \Delta^{\rm fluc} \varphi |_{({\bf x}, t;\, {\bf x'}, t')}
\, ,
\ee
where the r.m.s $\Delta^{\rm fluc}\varphi$ between two events (${\bf x}, t$) and (${\bf x'}, t'$) 
is defined as follows
\begin{eqnarray} \nonumber
\Delta^{\rm fluc} \varphi |_{({\bf x}, t;\, {\bf x'}, t')} ^2 \, 
& \equiv & \, \langle\, [\delta\varphi({\bf x}, t) - \delta\varphi({\bf x'}, t')]^2 \rangle  \\[2mm]
& = & \, \frac{\whh_{\rm ex}^2}{(2\pi)^3} \int \frac{d^3 k}{2\, k^3}\, 
\left[f_z(k)^2 + f_{z'}(k)^2
- 2 f_z(k) f_{z'}(k) \, e^{i {\bf k}({\bf x} - {\bf x'})} \right]\\[2mm]
& = & \, \frac{\whh_{\rm ex}^2}{(2\pi)^2} \int_0^\infty \frac{d k}{k}\, 
\left\{\left[f_z(k) - f_{z'}(k)\right]^2 + 2
f_z(k) f_{z'}(k)\left[1- \, \frac{\sin k x} 
{k x}\right]\right\}\, . \label{eq3.30}
\end{eqnarray}
Here, $x \equiv |{\bf x} - {\bf x'}|$ is the coordinate distance between the two events
and, consistently with the slow-roll approximation, the Hubble expansion rate at
exit has been assumed to be scale-invariant:
 $\whh_{\rm ex}(k) \simeq \whh_{\rm ex}  \simeq 3 \times 10^{-5}$ . 

If one considers spatial fluctuations over terrestrial or 
solar system proper length scales $l=a_0k^{-1} \ll H_0^{-1}$ at the present time 
$t=t'= t_0$, the first square brackets in (\ref{eq3.30}) vanishes and one can expand the 
sine function at small $kx$  
obtaining 
\be \label{eq3.28}
\Delta^{\rm fluc} \varphi |_{l;\, z=0} \, \simeq  \frac{\whh_{\rm ex}}{2\pi} \, 
\, \frac{H_0\, l}{\sqrt{3}} \qquad ; \, 
\left. \frac{\Delta^{\rm fluc} e^2}{e^2}\right|_{l;\, z=0} \, \simeq \, 10^{-2} \, 
\alpha_{\rm had}\, \whh_{\rm ex}\, H_0 \, l \, .
\ee
As expected, these variations are
extremely small, ${\Delta^{\rm fluc} e^2}/{e^2}|_{l;\, z=0}\simeq 10^{-33} \, 
l / {\rm km}$. 
It is also interesting to compare dilaton fluctuations at different redshifts 
along a comoving observer wordline. By putting $x\simeq 0$ in (\ref{eq3.30}) 
the second term in the square brackets vanishes and one has:
\be \label{eq3.34}
\Delta^{\rm fluc} \varphi |_{z;\, x=0} \, \simeq \, \frac{\whh_{\rm ex}}{2\pi}
\frac{1}{\sqrt{2}}
\left[\log(1+z) - \frac{z}{1+z} + \frac{10^{-8}}{2} z^2\right]^{1/2} \simeq 
\frac{\whh_{\rm ex}}{2\pi} \left[\frac{z}{2} - \frac{z^2}{3} + \dots\right] \, .
\ee

It is slightly  more complicated  to compare dilaton fluctuations
 between ``now'' and events 
at redshift $z$ along a null ray. Expanding in powers of $z$ around $z=0$  
one gets from (\ref{eq3.30}), after a straightforward calculation:
\be \label{eq3.29} 
\Delta^{\rm fluc} \varphi |_z  \,  \simeq \, \frac{\whh_{\rm ex}}{2\pi}\, 
\left[\frac{1}{2}\sqrt{\frac{7}{3}}\, z - \frac{2}{\sqrt{21}}\, z^2 + \dots \, \right] .
\ee
Numerically, at redshift $z \sim 1$, the effects of dilatonic fluctuations 
give  $\Delta^{\rm fluc}  \varphi |_{z=1} \sim \whh_{\rm ex}/(2 \pi) \sim 5 \times 10^{-6}$.
 This is to be
contrasted with the effects of the cosmic, 
homogeneous evolution which yields $\Delta  \varphi |_{z=1} \simeq \alpha_m$.
In the ``normal'' case where $\alpha_m \sim e^{-c \varphi} \sim \delta_H^{4/(n+2)}$,
the two effects, though a priori unrelated, are related  in our  scenario , when $n=2$.
Indeed, if $n=2$,  $\delta_H^{4/(n+2)} = \delta_H \sim 5 \times 10^{-5}$ is linked to
$\whh_{\rm ex}/(2 \pi)$ via $ \delta(\chi_{\rm ex}) = A \whh_{\rm ex}/(2 \pi)$
with $A = (8/3) V/ \partial_{\chi} V = (8/3) (\chi/n) \simeq 40/( 3 \sqrt{n}) \sim 10$.
On the other hand, in the case where $\varphi$ is strongly coupled to dark matter, the
homogeneous evolution $\Delta  \varphi |_{z=1} \simeq \alpha_m \sim 1$ is parametrically
larger than the fluctuations $\Delta^{\rm fluc}  \varphi |_{z=1} \sim \whh_{\rm ex}/(2 \pi)$.
 
To conclude on this subsection, we see that the inhomogeneous space-time fluctuations
of the fine-structure constant are typically too small to be observable (if the limits
from UFF are already satisfied), being suppressed, relative to their natural values 
$H_0 l\,, \, H_0 t$, by the  small factor $\alpha_{\rm had} \whh_{\rm ex} $.

\section{Summary and conclusion}
We have studied the dilaton-fixing mechanism of \cite{DP94} within the context where the 
dilaton-dependent low-energy couplings are extremized at $\varphi = + \infty$, i.e. for 
infinitely large values of the bare string coupling $g_s^2 = e^{\phi} \simeq e^{c \varphi}$. 
[The crucial coupling to the inflaton, say $\lambda(\varphi)$ in Eq.(\ref{eq2.10}), must be
{\it minimized} at $ \varphi \to + \infty$; the other couplings can be either minimized
or maximized there.]
This possibility of a fixed point at infinity (in bare string coupling space) has been 
recently suggested \cite{V01}, and its late-cosmological consequences have been explored in 
\cite{GPV}. We found that a primordial inflationary stage, with inflaton potential $V(\chi) = 
\lambda (\varphi) \, \chi^n / n$, was much less efficient in decoupling a dilaton with least 
couplings at infinity than in the case where the least couplings are reached at a finite value 
of $\varphi$ (as in \cite{DP94,DV96}). This reduced efficiency has interesting 
phenomenological consequences. Indeed, it predicts much larger observable deviations from 
general relativity. In the case of the simplest chaotic potential \cite{L90} $V(\chi) = 
\frac{1}{2} \, m_{\chi}^2 (\varphi) \, \chi^2$, we find that, under the simplest assumptions 
about the pre-inflationary state, this scenario predicts violations of the universality of 
free fall (UFF) of order $\Delta a / a \sim 5 \times 10^{-4} \, \delta_H^2$ where $\delta_H$ 
is the density fluctuation generated by inflation on horizon scales. The observed level of 
large-scale density (and cosmic microwave background temperature) fluctuations fixes 
$\delta_H$ to be around $5 \times 10^{-5}$ which finally leads to a prediction for a violation 
of the UFF near the $\Delta a / a \sim 10^{-12}$ level. This is naturally compatible with 
present experimental tests of the equivalence principle, and suggests that a modest 
improvement in the precision of UFF tests might be able to detect a deviation linked to 
dilaton exchange with a coupling reduced by the attraction towards the fixed point at 
infinity. Because of the presence of unknown dimensionless
 ratios $(c, b_F / b_\lambda)$ in 
our estimates, and of quantum noise in the evolution of the dilaton,
 we cannot give sharp quantitative estimates of $\Delta a / a$. However, we note 
that dilaton-induced violations of the UFF have a rather precise signature with a 
composition-dependence of the form (\ref{eq3.15}), with probable domination by the last 
(Coulomb energy) term \cite{DP94}. As explored in \cite{D96} this signature is quite distinct 
from UFF violations induced by other fields, such as a vector field. We note that the approved 
Centre National d'Etudes Spatiales (CNES) mission MICROSCOPE \cite{Touboul} (to fly in 
2004) will explore the level $\Delta a / a \sim 10^{-15}$, while the planned National 
Aeronautics and Space Agency (NASA) and European Space Agency (ESA) mission STEP (Satellite 
Test of the Equivalence Principle)\cite{worden}
 could explore the $\Delta a / a \sim 10^{-18}$ level. 
Our scenario gives additional motivation for such experiments and suggests that they might find 
a rather strong violation signal, whose composition-dependence might then be studied in detail 
to compare it with Eq.~(\ref{eq3.15}).

In the case of inflationary potentials $V(\chi) \propto \chi^n$ with $n > 2$ our simplest 
estimates predict a violation of the UFF of order $\Delta a / a \sim 5 \times 10^{-4} \, 
\delta_H^{\frac{8}{n+2}}$ which is larger than $10^{-12}$. At face value this suggests that 
existing UFF experimental data can be interpreted as favouring $n \leq 2$ over $n > 2$. 
However, we must remember that our estimates have made several simplifying assumptions. It is 
possible that the large quantum fluctuations of the inflaton in the self-regenerating regime 
$\chi > \chi_{\rm in}$, with $\chi_{\rm in}$ defined by Eq.~(\ref{eq2.19}), can give more time 
for $\varphi$ to run away towards large values, so that the effective value of $e^{c 
\varphi_{\rm in}}$ to be used in Eq.~(\ref{eq2.23}) turn out to dominate the first term in the 
R.H.S. that we have used for our estimates. We leave to future work a study of the system of
Langevin equations describing  the coupled fluctuations of $\phi$ and $\chi$ during the
self-regenerating regime.

Finally let us note some other conclusions of our work.

  We recover the conclusion of previous 
works on dilaton models that the most interesting experimental probes of a massless weakly 
coupled dilaton are tests of the UFF. The composition-independent gravitational tests (solar-system, 
binary-pulsar) tend to be much 
less sensitive probes (as highlighted by the relations (\ref{eq3.18}), (\ref{eq3.23}) and 
(\ref{eq3.24})).

 However, a possible exception concerns the time-variation of the
coupling constants. Here the conclusion depends crucially on the assumptions made about the
couplings of the dilaton to the cosmologically dominant forms of energy (dark matter and/or
dark energy). If these couplings are of order unity (and as large as is phenomenologically 
acceptable, i.e. so that $(\varphi'_0)^2 = 0.7$),
the present time variation of the fine-structure constant is linked to the violation
of the UFF by the relation $d \ln e^2 / dt \sim \pm 2.0 \times 10^{-16}
\,   \sqrt{ 10^{12} \Delta a/a} {\rm yr}^{-1}$. [The most natural sign here being $+$, 
i.e. $b_F > 0$, which corresponds to {\it smaller} $e^2$ in the past, 
just as suggested by the claim \cite{Webb}.]
Such a time variation might be observable
( if $\Delta a/a$ is not very much below its present upper bound $\sim 10^{-12}$) through
the comparison of high-accuracy cold-atom clocks and/or via improved measurements of
astronomical spectra.

 More theoretical work is needed to justify the basic assumption 
(\ref{eq1.3}) of our scenario. In particular, it is crucial to investigate whether it is 
natural to expect that the sign of the crucial coefficient $b_{\lambda}$ in Eq.~(\ref{eq2.11}) 
be indeed {\it positive}. [Recall that the general mechanism of \cite{DP94} is an attraction 
towards ``Least Couplings'' while Eq.~(\ref{eq1.3}) with ${\cal O}(e^{- \phi}) > 0$ leads
to largest couplings at infinity.] Note in this respect that the sign of the other $b_i$'s is not 
important as, once inflation has pushed $e^{c \varphi}$ to very large values $e^{c 
\varphi_{\rm end}}$, the subsequent cosmological evolutions tend to be ineffective in further 
displacing $\varphi$. 
\acknowledgments
It is a pleasure to thank Ian Kogan for suggesting a method to solve the (decoupled) Langevin
equation for $\varphi$.
GV   wishes to
acknowledge the support of a ``Chaire Internationale Blaise
Pascal", administered by the ``Fondation de L'Ecole Normale
Sup\'erieure''. The work of F.P. was supported in part
by INFN and MURST under contract 2001-025492, and by the European
Commission TMR program HPRN-CT-2000-00131, in association to the
University of Padova."

\appendix 

\section{The stochastic evolution of the dilaton}

In this appendix we study the stochastic evolution of the dilaton $\varphi$ during 
inflation as described by the Langevin-type equation (\ref{eq4.2}). 
We restrict our attention to the region of 
phase space where the evolution of the inflaton
$\chi$ is classical, and to a power-law potential of the form (\ref{eq2.10}). 
It follows that the inflaton evolves according to the classical slow-roll equation 
(\ref{eq2.13}) whose solution reads
\be \label{a0} 
\chi^2 = \chi_{\rm in}^2 - n p,
\ee 
where $p$, the 
parameter defined in (\ref{eq2.7}), is shifted in such a way that $p_{\rm in} \equiv 0$. 
Equation  (\ref{eq4.2}) takes the form 
\begin{equation} \label{a1}
\frac{d \varphi}{dp} = \frac{1}{2} \, b_{\lambda} \, c \, e^{-c\varphi} +  \,
\xi (p) \, ,
\end{equation} 
where $\xi(p)$ is a gaussian stochastic variable (GSV), with a ``time-dependent'' r.m.s. 
amplitude $\whh (p) /2\pi$:
\be \label{a3}
\langle\xi(p_1)\, \xi(p_2) \rangle \, \, =\, \frac{\whh^2}{(2\pi)^2}\, \delta(p_1-p_2), 
\ee
[the relation with the normalized random white noise term of (\ref{eq4.2}) is
$\xi(p) = \xi_2(p) \whh / 2\pi$].
For any given source term $\xi(p)$, the formal solution of (\ref{a1}) reads
\be \label{a4}
e^{c\, \varphi(p)} \, =\, e^{c\, \varphi_{\rm in}} \, e^{c\, \eta(p)} + 
\frac{b_\lambda c^2}{2}
\int_0^p dp' e^{c\, [\eta(p) - \eta(p')]}\, ,
\qquad
\eta(p) \equiv \int_0^p dp'\, \xi(p')\, . 
\ee
Note that the classical solution in (\ref{eq2.16}), 
$e^{c\, \varphi_{\rm cl}(p)} = e^{c\, \varphi_{\rm in}} + (b_\lambda c^2 /2)\,  p$ ,
can be easily recovered in the 
small noise limit $\xi(p) \rightarrow 0$, $\eta(p) \rightarrow 0$. 

It proves convenient to compare the true solution
 to the classical one by studying the statistical 
behaviour of the ratio $A(p) \equiv e^{c\, \varphi(p)}/ e^{c\, \varphi_{\rm cl}(p)}$.
As we will show below, $\langle e^{c\, \eta(p)} \rangle  = {\cal O}(1)$. Moreover, we 
are also assuming $e^{c\, \varphi_{\rm in}} = {\cal O}(1)$ or, at least,
$e^{c\, \varphi_{\rm in}} \ll (b_\lambda c^2/2) p$ (see Section II for details)
so that the leading contribution to the first equation in (\ref{a4}) is given by the integral, 
and we have
\be\label{a4'}
A(p) \, \equiv \, e^{c\, \varphi(p)}/ e^{c\, \varphi_{\rm cl}(p)} \, \simeq \, 
\frac{1}{p} 
\int_0^p dp' e^{c\, [\eta(p) - \eta(p')]}\, .
\ee 
Since $\xi(p)$ is a GSV, also its integral $\eta(p)$ is a (centered) GSV. Moreover,
if $x$ is a GSV with $\sigma_x^2 \, \equiv \, \langle x^2 \rangle  - \langle x \rangle ^2 \, = \, \langle x^2 \rangle $, 
by Bloch's theorem $y= e^x$ is a new stochastic variable with 
$\langle y \rangle  \, = \, \langle e^x \rangle  \, = \, e^{\langle x^2 \rangle /2}$ and 
$\sigma_y^2 \, = \, e^{2 \langle x^2 \rangle } - e^{\langle x^2 \rangle }$.
The average value of $A(p)$ thus reads 
\be 
\label{a5}
\langle A(p) \rangle   \, \, \simeq \,
\frac{1}{p} \int_0^p dp' \, e^{(c^2/2)\, \langle[\eta(p) - \eta(p')]^2 \rangle }\,. 
\ee
The exponent on the right hand side of the above equation can be estimated by 
using (\ref{a3}) and the slow-roll approximation  
$\whh^2 \simeq 2 V(\chi,\varphi)/3 = 2 \lambda(\varphi) \chi^n /3n \simeq 2 
\lambda_\infty \chi^n/3n $. One gets
\be \label{a6}
\langle[\eta(p)-\eta(p')]^2 \rangle \, \, =\, 
\frac{1}{(2\pi)^2} \int_{p'}^{p} \whh^2 dp''\, \simeq \, \frac{n}{2(n+2)}
\left[\left(\frac{\chi(p')}{\chi_{\rm in}}\right)^{n+2}\!\!\! -\left( 
\frac{\chi(p)}{\chi_{\rm in}}\right)^{n+2} \right]\, ,
\ee
where $\chi_{\rm in}$ is the  value at exit from self-regenerating inflation: 
$\whh(\chi_{\rm in})/2\pi = n /(2\chi_{\rm in})$ (see Section II for more details).
Since we are interested in evaluating (\ref{a5}) at the end of inflation, 
$p = p_{\rm end} \simeq \chi_{\rm in}^2/n$, we can thus write
\be \label{a8}
\langle[\eta(p)-\eta(p')]^2 \rangle \, \, \simeq\, 
\frac{n}{2(n+2)} \left[1-\frac{p'}{p_{\rm end}}\, \right]^{(n+2)/2}.
\ee
When evaluated at $p'=0$, the above formula gives  
$\langle\eta(p)^2 \rangle  = n/(2(n+2))$. Thus the normalization factor to the initial condition in 
(\ref{a4}) is of order one, as anticipated: $\langle e^{c\, \eta(p)} \rangle  =  e^{ \frac{1}{2} c \langle 
\eta(p)^2 \rangle} = {\cal O}(1)$.
{}From (\ref{a5}) and (\ref{a8}) we have:
\begin{eqnarray} \nonumber
\langle A(p_{\rm end})  \rangle  \, 
&\simeq &\, \,  \frac{1}{p_{\rm end}} 
\int_0^{p_{\rm end}} dp' \exp\left[\frac{c^2 n}{4(n+2)}\left(1-\frac{p'}{p_{\rm end}}
\right)^{(n+2)/2}\right]   \\
& = & \, 
\int_0^1 \, \exp\left[\frac{c^2 n}{4(n+2)} 
x^{(n+2)/2}\right] dx  \\ \nonumber
& = & \,  \exp\left(\frac{c^2 n \theta}{4(n+2)}\right) = {\cal O}(1) ,
\end{eqnarray}
with $0 < \theta < 1$.

 We can estimate the dispersion of the same quantity by  expanding
the exponential inside the integral (\ref{a5}) 
in powers of $\xi(p)$:
\be
A(p) \, \simeq \, 
1 + \frac{c}{p} \int_0^p dp' \int_{p'}^p dp'' \xi(p'') + \frac{1}{2} \frac{c^2}{p} 
\int_0^p dp'\left(\int_{p'}^p dp''\xi(p'')\right)^2 + \dots .
\ee
At lowest order in $\xi(p)$ the variance of the above quantity calculated at 
$p = p_{\rm end}$ reads
\begin{eqnarray} \nonumber
\sigma^2_{A(p_{\rm end})}\,  & = &
\left\langle \left(\frac{c}{p} \int_0^p dp' \int_{p'}^p dp'' \xi(p'')\right)^2\right \rangle \\
&=& \frac{c^2}{p^2}  \int_0^p dp'  \int_0^p dp''\int_{{\rm max}(p',p'')}^p dp''' 
\frac{\whh^2(p''')}{(2\pi)^2} \\ \nonumber
&=& \frac{2 c^2}{p^2} \int_0^p dp' p' \int_{p'}^p dp'' \frac{\whh^2(p'')}{(2\pi)^2} .
\end{eqnarray}
As in equation (\ref{a6}) we can use the slow-roll approximation and obtain 
\be
\sigma^2_{A(p_{\rm end})}\  =  \
\frac{c^2\, n}{n+2}\int_0^1 x\, (1-x)^{(n+2)/2} dx\ 
= \ c^2 {\cal O}(1) \, . \nonumber
\ee

\end{document}